\documentclass[preprint,showpacs,preprintnumbers,amsmath,amssymb,natbib]{revtex4}

\usepackage{graphicx}
\usepackage{epsfig}		
\usepackage{dcolumn}
\usepackage{bm}
\usepackage{amsmath}

\def\0{\mbox{\tiny $0$}}
\def\1{\mbox{\tiny $1$}}
\def\2{\mbox{\tiny $2$}}
\def\3{\mbox{\tiny $3$}}
\def\4{\mbox{\tiny $4$}}
\def\5{\mbox{\tiny $5$}}
\def\6{\mbox{\tiny $6$}}
\def\7{\mbox{\tiny $7$}}
\def\8{\mbox{\tiny $8$}}
\def\9{\mbox{\tiny $9$}}

\def\f14{\mbox{\tiny $\frac{1}{4}$}}

\def\L{\mbox{\tiny $L$}}

\def\R{\mbox{\tiny $R$}}

\def\mi{\mbox{\tiny $-$}}

\def\bb#1{\mbox{\footnotesize $(#1)$}}
\def\mt#1{\mbox{\textsl{#1}}}
\def\mbf#1{\mbox{\boldmath$#1$}}
\def\bb#1{\mbox{\small $(#1)$}}

\DeclareMathOperator{\Tr}{\mbox{Tr}}

\begin{document}

\title{Quantum transitions and quantum entanglement from Dirac-like dynamics simulated by trapped ions}

\author{Victor A. S. V. Bittencourt}
\email{vbittencourt@df.ufscar.br}
\author{Alex E. Bernardini}
\email{alexeb@ufscar.br}
\affiliation{Departamento de F\'{\i}sica, Universidade Federal de S\~ao Carlos, PO Box 676, 13565-905, S\~ao Carlos, SP, Brasil}
\author{Massimo Blasone}
\email{blasone@sa.infn.it}
\affiliation{Dipartimento di Fisica, Universit\`a Di Salerno, Via Giovanni Paolo II, 132 84084 Fisciano, Italy}
\altaffiliation{Also at: INFN Sezione di Napoli, Gruppo collegato di Salerno, Italy}

\begin{abstract}
Quantum transition probabilities and quantum entanglement for two-{\em qubit} states of a four level trapped ion quantum system are computed for time-evolving ionic states driven by Jaynes-Cummings Hamiltonians with interactions mapped onto a $\mbox{SU}(2)\otimes \mbox{SU}(2)$ group structure.
Using the correspondence of the method of simulating a $3+1$ dimensional Dirac-like Hamiltonian for bi-spinor particles into a single trapped ion, one preliminarily obtains the analytical tools for describing ionic state transition probabilities as a typical quantum oscillation feature.
For Dirac-like structures driven by generalized Poincar\'e classes of coupling potentials, one also identifies the
$\mbox{SU}(2)\otimes \mbox{SU}(2)$ internal degrees of freedom corresponding to {\em intrinsic parity} and {\em spin polarization} as an adaptive platform for computing the quantum entanglement between the internal quantum subsystems which define two-{\em qubit} ionic states.
The obtained quantum correlational content is then translated into the quantum entanglement of two-{\em qubit} ionic states with quantum numbers related to the total angular momentum and to its projection onto the direction of the trapping magnetic field.
Experimentally, the controllable parameters simulated by ion traps can be mapped into a Dirac-like system in the presence of an electrostatic field which, in this case, is associated to ionic carrier interactions.
Besides exhibiting a complete analytical profile for ionic quantum transitions and quantum entanglement, our results indicate that carrier interactions actively drive an overall suppression of the quantum entanglement.
\end{abstract}

\date{\today}
\maketitle

\section{Introduction}

The experimental engineering of trapped ion platforms adapted for detecting local quantum correlations, simulating open-system dynamical maps, building microwave quantum logic gates, and measuring quantum phase transitions \cite{Nat01,Nat02,Nat03,Nat04,Nat05} have raised the state-of-the-art in producing nanotechnologies up to a novel and challenging baseline.

Since it simulates several quantum effects as they were driven by a Dirac-like Hamiltonian \cite{n001,n002,n003,Nat06}, on the theoretical front, the trapped ion physics has also worked as a convenient operational tool for testing the interface between the relativistic quantum mechanics and the solid state physics.
Through the map of a Jaynes-Cummings Hamiltonian dynamics onto a $\mbox{SU}(2)\otimes \mbox{SU}(2)$ group structure, the ion-trap technology has provided a novel routine to phenomenologically access and manipulate the interface between the trapped ion physics and the relativistic quantum mechanics of the Dirac equation \cite{n004,n005,n006}.
For instance, trapped ion interacting Hamiltonians once mapped onto the structure of the Dirac equation can straightforwardly reproduce typical quantum effects of relativistic nature, such as the \textit{zitterbewegung}/trembling motion \cite{n001}, the Klein paradox \cite{n002}, or even the spinor-motion correlation inherent to the tachyonic dynamics \cite{new01}.
Moreover, quantum correlations between $\mbox{SU}(2)\otimes \mbox{SU}(2)$ internal degrees of freedom of {\em intrinsic parity} and {\em spin polarization} of Dirac particles \cite{n008,n009} (corresponding to $\mbox{SU}(2)\otimes \mbox{SU}(2)$ bi-spinors) can work as an efficient quantifier of two-{\em qubit} entanglement of trapped ion structures.

The creation and manipulation of entanglement through Jaynes-Cummings interactions can also be relevant for the implementation of quantum algorithms \cite{entjc01} and for the characterization of classical to quantum transitions \cite{entjc04}. For instance, when a Dirac oscillator \cite{entjc02} is investigated, the entanglement between intrinsic degrees of freedom of the bi-spinor and its orbital angular momentum is identified by the entanglement between discrete levels of the ion and their vibrational degrees of freedom \cite{entjc03}, which supports some signatures of chiral quantum phase transitions \cite{n014}.

Considering that the Dirac Hamiltonian may be written in terms of the direct product of two-{\em qubit} operators \cite{n008,n009}, the $\mbox{SU}(2)\otimes \mbox{SU}(2)$ group structure (c. f. the Appendix) involving such \textit{spin-parity} internal degrees of freedom exhibits an energy spectrum associated to two-{\em qubit} quantum correlated states \cite{n008}.
In particular, the inclusion of additional Dirac-like global potentials driven by (pseudo)scalar, (pseudo)vector and (pseudo)tensor interactions can also create novel patterns of intrinsic $\mbox{SU}(2)\otimes \mbox{SU}(2)$ quantum correlations as well as destroy the separability of an eventual free particle state \cite{n009}.

The complete overview of the entanglement driven by Poincar\'e classes of $\mbox{SU}(2) \otimes \mbox{SU}(2)$ coupling potentials \cite{n009} can be specialized for more feasible ionic systems as, for instance, those ones that simulate the behavior of the electric dipole moment of a spin one-half particle in an electromagnetic field \cite{n005,n010}.
In this case, the Hamiltonian for a neutral Dirac particle with momentum $\bm{p}$ and mass $m$ non-minimally coupled to external electric and magnetic fields, $\bm{\mathcal{E}}$ and $\bm{\mathcal{B}}$, is given by
\begin{equation}
\label{eqsA0}
\hat{\mathcal{H}} = c \,\hat{\bm{\alpha}} \cdot \bm{p} + \hat{\beta}\, m c^2+ \kappa \,\hat{\beta}\, ( \,\hat{\bm{\Sigma}} \cdot \bm{\mathcal{E}} +i\, c\,\hat{\bm{\alpha}} \cdot \bm{\mathcal{B}} \,)\, + \mu\, \hat{\beta}\,( i\, \hat{\bm{\alpha}} \cdot \frac{\bm{\mathcal{E}}}{c} - \hat{\bm{\Sigma}} \cdot \bm{\mathcal{B}}\,),
\end{equation}
where $\kappa$ and $\mu$ are respectively the electric and magnetic dipole moments, with {\em boldface} variables used to denote vectors, $\bm{a}$, with $ a = \vert \, \bm{a} \, \vert = \sqrt{\bm{a}\cdot\bm{a}}$, and {\em hat} ``$~\hat{}~$'' used to denote Dirac operators.
In the above Hamiltonian, $\hat{\beta}$ and $\hat{\bm{\alpha}} \equiv ( \hat{\alpha}_1, \hat{\alpha}_2, \hat{\alpha}_3)$ are the Dirac matrices that must satisfy the anti-commuting relations
$\{\hat{\alpha}_i, \hat{\alpha}_j\} = 2 \, \delta_{ij} \, \hat{I}_4$, and
$\{\hat{\alpha}_i, \hat{\beta}\} =0$, with $i,j = 1,2,3$, and
$\hat{\beta}^2 = \hat{I}_4$ (where $\hat{I}_N$ denotes the $N\times N$ identity operator).
Assuming that Dirac matrices are expressed through different representations interconnected by unitary transformations, one can consider a particular representation given by
\begin{equation}
\hat{\bm{\alpha}} = \hat{\sigma}_x \otimes \hat{\bm{\sigma}} \equiv \left[ \begin{array}{rr} 0 & \hat{\bm{\sigma}} \\ \hat{\bm{\sigma}} & 0 \end{array}\right],
\qquad \mbox{and} \qquad
\hat{\beta} = \hat{\sigma}_z \otimes \hat{I}_2 \equiv \left[ \begin{array}{rr} \hat{I}_2 & 0 \\ 0 & - \hat{I}_2 \end{array} \right],
\label{eqsAAA}
\end{equation}
where ${\bm{\sigma}}$ are the Pauli matrices, and one identifies the matrices $\hat{\bm{\Sigma}} = \hat{I}_2 \otimes \hat{\bm{\sigma}}$ as those related to the spin operator given by $\hat{\bm{S}} = {\hat{\bm{\Sigma}}}/{2}$.
For a neutral particle moving on an electrostatic field with respect to the laboratory frame, $\mathcal{\bm{B}}=0$ and the Hamiltonian from ($\ref{eqsA0}$) can be simplified into
\begin{equation}
\label{eqsA00}
\hat{\mathcal{H}} = c\, \hat{\bm{\alpha}} \cdot \bm{p} + \hat{\beta}\, m c^2+ \kappa \,\hat{\beta}\, \hat{\bm{\Sigma}} \cdot \bm{\mathcal{E}} \, + i \mu\, \hat{\beta}\, \hat{\bm{\alpha}} \cdot \frac{\bm{\mathcal{E}}}{c},\end{equation}
which can be effectively simulated by Jaynes-Cummings and carrier interactions between internal ionic states \cite{n002}.
The above dynamics can be used to compute the quantum correlational content of the corresponding two-{\em qubit} ionic states.

Given the above-mentioned baselines, the main purpose of this work is concerned with identifying and quantifying the entanglement and quantum correlations of internal ionic states encoded by a Dirac-like bi-spinor structure. The Hamiltonian dynamics from Eq.~(\ref{eqsA00}) can be reproduced by a suitable trapped ion setup \cite{n005} such that its eigenstates are given in terms of a superposition of four internal ionic states, which spans the four dimensional Hilbert space associated to the Dirac bi-spinor discrete degrees of freedom.
By suitably mapping the ionic state basis onto the complete set of four eigenstates of (\ref{eqsA00}) \cite{n009}, the dynamics of the ionic states can be entirely described by the Dirac $\mbox{SU}(2) \otimes \mbox{SU}(2)$ structure.
Then, the transition probabilities between the different internal ionic levels (once driven by the Dirac-like dynamics), and the entanglement/separability between states with different angular momenta can be straightforwardly computed, as well as their origins can be identified in terms of Dirac-like observables related to the encoded quantum concurrence between {\em spin polarization} and {\em intrinsic parity}.
In addition, a connection between the averaged chirality \cite{n011,n011A,n011B} defined as the average value of the operator $\hat{\gamma}_5 = - i \hat{\alpha}_x \, \hat{\alpha}_y \, \hat{\alpha}_z$, and the quantum concurrence between {\em spin polarization} and {\em intrinsic parity} can be evaluated and re-interpreted in terms of ionic state observables.
The averaged chirality can also be identified as a measurement of the maximal superposition between two of the four internal ionic levels.
To summarize, a complete prospect of quantum transitions and quantum entanglement, via quantum concurrence, for trapped ion systems driven by Jaynes-Cummings interactions can be mapped and computed in terms of the $\mbox{SU}(2) \otimes \mbox{SU}(2)$ Dirac-like structure, similar as it has been performed in \cite{n009,n010}.
By construction, the $\mbox{SU}(2)\otimes \mbox{SU}(2)$ {\em spin-parity} quantum correlational content can be interpreted in terms of the quantum entanglement of two-{\em qubit} ionic states for which the quantum numbers are related to the total angular momentum and to its projection onto the direction of the trapping magnetic field.

The manuscript is therefore organized as following.
In Sec. II, the correspondence between trapped ion Jaynes-Cummings interactions and some particular Poincar\'e classes of $\mbox{SU}(2) \otimes \mbox{SU}(2)$ coupling potentials is identified.
The corresponding trapped ion state parameters are mapped into a Dirac-like system in the presence of an electric field, as driven by Eq.~(\ref{eqsA00}).
In Sec. III, the Dirac-like eigenstates of (\ref{eqsA00}) are obtained, and the bi-spinor entangling properties are preliminarily discussed.
The connection between chirality and measurements of a maximal superposition between internal ionic states is also identified.
The main results of the paper are obtained along Sec. IV. The transition probabilities between the internal levels are calculated, an internal ionic dynamics is recovered, and the corresponding intrinsic (Dirac-driven) quantum concurrence as an entanglement quantifier for each time-evolving ionic state is obtained and confronted with the quantum transition profile.
Our final conclusions are drawn in Sec. V.

\section{Jaynes-Cummings Hamiltonian interactions mapped onto the $\mbox{SU}(2)\otimes \mbox{SU}(2)$ group structure}

\begin{figure}
\includegraphics[width = 8.0 cm]{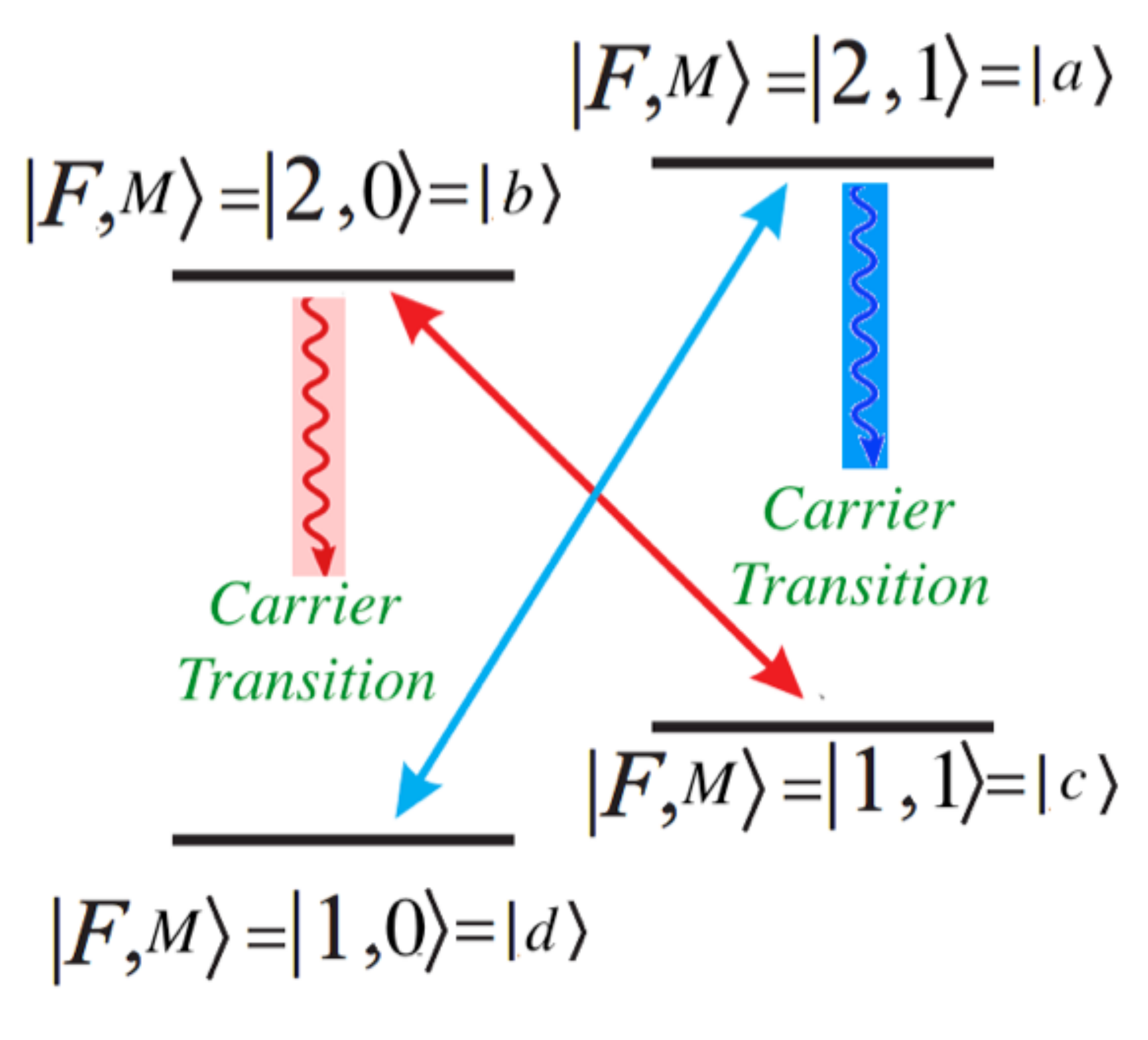}
\caption{Pragmatic scheme for the hyperfine levels and corresponding JC (red-sideband) and AJC (blue-sideband) transitions from ground states of the alkali ions, such as Mg$^+$, Ca$^+$, Sr$^+$ and Ba$^+$. The energy levels are identified by atomic labels $\vert F, M\rangle$, where $F$ is the quantum number for total angular momentum and $M$ is the analogous for the projection of the angular momentum onto the trap magnetic field direction. Such a configuration suggests the {\em qubit} assignment introduced by (\ref{eqsA08}).}
\label{eqsfig:level}
\end{figure}

The simulation of the Dirac Hamiltonian dynamics requires the confinement of an ion of mass $\tilde{m}$ by an electromagnetic trap, as for instance, through a radio frequency potential in a Paul trap \cite{n012}.
The ion oscillates with frequencies $\nu_x, \, \nu_y, \, \nu_z$ along the directions $x, \,y, \, z$ such that four metastable internal ionic states, $\{\vert a \rangle, \, \vert b \rangle, \, \vert c \rangle, \, \vert d \rangle \}$ are coupled pairwise with the ionic motion by an auxiliary electromagnetic field.
For a strongly confined ion engendered by a suitable tuning between the driven electromagnetic field and the trapping potential, such a coupling between internal states and the ion motion is described, in the rotating wave approximation (i. e. by neglecting rapidly oscillating terms) by the Jaynes-Cummings (JC) and the anti-Jaynes-Cummings (AJC) interactions, respectively corresponding to red-sideband and blue-sideband excitations, through the Hamiltonians \cite{n012}
\begin{equation}
\label{eqsA01}
\hat{H}_j ^{JC} = \hbar \eta_j \tilde{\Omega}_j \, (\,\hat{\sigma}^+ a_j e^{i \phi_r} + \hat{\sigma}^- a_j^\dagger e^{-i \phi_r} \,) + \hbar \delta_j \hat{\sigma}_z,
\end{equation}
and
\begin{equation}
\label{eqsA02}
\hat{H}_j ^{AJC} = \hbar \eta_j \tilde{\Omega}_j \, (\,\hat{\sigma}^+ a_j^\dagger e^{i \phi_b} + \hat{\sigma}^- a_j e^{-i \phi_b} \,) + \hbar \delta_j \hat{\sigma}_z,
\end{equation}
with $j = x,\,y,\,z$, where $\phi_{r\,(b)}$ are the red(blue)-sideband phases, $\tilde{\Omega}_j$ are the Rabi frequencies, $\hat{\sigma}^{+ \, (-)}$ are the raising and lowering ladder operators between the corresponding two internal levels, and $\eta_j = k \sqrt{\hbar/2 \tilde{m} \nu_j}$ is the Lamb-Dicke parameter (where $k$ is the wave number of the driving field and $\tilde{m}$ is the ion mass). The parameter $\delta_j$ is called the detuning (frequency) between the field and the two-level system.
The JC interaction excites the vibrational level while de-excites the internal state. On the other hand, the AJC interaction promotes the excitation of both vibrational and internal levels. A pictorial scheme for such interactions is shown in Fig.~\ref{eqsfig:level}.
A third interaction that arises when one considers an ion in the above trapping regime is the carrier interaction given by the Hamiltonian \cite{n012}
\begin{equation}
\label{eqsA03}
\hat{H}_j^C = \hbar \Omega_j (\hat{\sigma}^+ e^{i \phi} + \hat{\sigma}^- e^{-i \phi}),
\end{equation}
which accomplishes an excitation of the internal levels and does not change the vibrational state of the ion. The three interactions - JC, AJC and carrier ones - are resonances of an interaction Hamiltonian that describes the coupling between the external electromagnetic field and the trapped ion when one considers the regime where the ion wave function extension is much smaller than $1/k$, that is, the so called Lamb-Dicke regime \cite{n012}.

By suitable choices of the driving phases, the combination of the above introduced three interactions reproduces the dynamics of a Dirac Hamiltonian including external fields \cite{n001,n002,n003,n004,n005}. Depending on the dimension of the subjacent space-time, on the representation of Dirac matrices and on the interacting external fields,
a particular setup can be used to engender the Dirac equation dynamics.
To map the dynamics driven by a non-minimally coupling with an electric field, described by the Hamiltonian (\ref{eqsA00}), the procedure introduced by \cite{n005} can be straightforwardly evaluated.
One firstly notices that the Dirac mass term, $\hat{\beta}\, m c^{\2}$, can be mapped into
\begin{equation}
\label{eqsA04}
\hat{\beta} m c^2 \rightarrow 2 \hbar \delta( \hat{\sigma}_z^{ad} + \hat{\sigma}_z ^{bc}),
\end{equation}
where the upper script index denotes the internal levels involved.
As an example, for $\hat{\sigma}_z^{ad}$ and $\hat{\sigma}_z^{bc}$, one has
\begin{equation}
\label{eqssigs}\hat{\sigma}_z^{ad} \equiv \vert a \rangle \langle a \vert - \vert d \rangle \langle d \vert= \left[ \begin{array}{lrrr} 1 & 0 & 0 & 0 \\ 0 & 0 & 0 & 0 \\ 0 & 0 & 0 & 0 \\ 0 &\phantom{-} 0 & \phantom{-}0 & -1 \end{array} \right],\quad
\hat{\sigma}_z^{bc} \equiv \vert b \rangle \langle b \vert - \vert c \rangle \langle c \vert= \left[ \begin{array}{lrrr} 0 & 0 & 0 & 0 \\ 0 & 1 & 0 & 0 \\ 0 &\phantom{-} 0 & -1 & \phantom{-}0\\0 & 0 & 0 & 0 \end{array} \right],
\end{equation}
such that $\hat{\sigma}_z^{ad} + \hat{\sigma}_z ^{bc} \equiv \hat{\beta}$.
Analogously, the momentum term $c \,\hat{\bm{\alpha}} \cdot \bm{p}$ can be reproduced by
\begin{equation}
\label{eqsA05}
c \, \hat{\bm{\alpha}}\cdot \bm{p} \rightarrow 2 \eta \Delta_x \tilde{\Omega}(\hat{\sigma}_x^{ad} + \hat{\sigma}_x^{bc}) p_x + 2 \eta \Delta_y \tilde{\Omega}(\hat{\sigma}_y^{ad} - \hat{\sigma}_y^{bc})p_y + 2 \eta \Delta_z \tilde{\Omega}(\hat{\sigma}_x^{ac} - \hat{\sigma}_x^{bd}) p_z,
\end{equation}
where, in the same sense of (\ref{eqssigs}), for $\hat{\sigma}_x^{ad}$ and $\hat{\sigma}_x^{bc}$, one has
\begin{equation}
\hat{\sigma}_x^{ad} \equiv \vert a \rangle \langle d \vert + \vert d \rangle \langle a \vert= \left[ \begin{array}{rrrr} ~0 & ~0 & ~0 & ~1 \\ 0 & 0 & 0 & 0 \\ 0 & 0 & 0 & 0 \\ 1 & 0 & 0 & 0 \end{array} \right],\quad \hat{\sigma}_x^{bc} \equiv \vert b \rangle \langle c \vert + \vert c \rangle \langle b \vert= \left[ \begin{array}{rrrr} ~0 & ~0 & ~0 & ~0 \\ 0 & 0 & 1 & 0 \\ 0 & 1 & 0 & 0 \\ 0 & 0 & 0 & 0 \end{array} \right],
\label{eqsBBB}
\end{equation}
with analogous notation for the additional interactions.
The $j$-th component of the momentum is given in terms of the vibrational state of the ion by the map
\begin{equation}
p_j \rightarrow \frac{i \hbar}{2 \Delta_j} \, (\,a_j ^\dagger - a_j \,),
\end{equation}
where $\Delta_j = \sqrt{\hbar/ 2 \tilde{m} \nu_j}$ is the position spreading of the ion ground state wave function.
For instance, to simulate the $p_x$ term of Dirac equation one might choose $\phi_r = - \pi/2$ and $\phi_b = \pi/2$ into Eqs.~(\ref{eqsA01}) and (\ref{eqsA02}). By requiring space homogeneity, the frequency parameters are constrained by $\nu_x = \nu_y = \nu_z = \nu$, consequently $\tilde{\Omega}_j = \tilde{\Omega}$, $\Delta_j = \Delta$ and $\eta_j = \eta$ for all directions ($j = x,\,y,\,z$), such that the {\em free particle terms} of the Dirac equation $c \, \bm{p} \cdot \hat{\bm{\alpha}} + \hat{\beta}\,m c^2$ shall be reproduced by the sum of the JC and the AJC interactions, Eqs.~(\ref{eqsA04}) and (\ref{eqsA05}), respectively.

Through a convenient choice of the phase $\phi$, the {\em tensor and pseudotensor potential terms}, $\kappa \,\hat{\beta}\, \hat{\bm{\Sigma}} \cdot \bm{\mathcal{E}}$ and $i \mu\, \hat{\beta}\, \hat{\bm{\alpha}} \cdot \bm{\mathcal{E}}/{c}$, are then mapped through two carrier interactions (\ref{eqsA03}) with frequencies $\Omega_j^{(1)}$ and $\Omega_j^{(2)}$:
\begin{subequations}
\label{eqsA06}
\begin{equation}
\label{eqsA06A}
\hat{\beta} \hat{\bm{\Sigma}} \cdot \left( \kappa \, \bm{\mathcal{E}} \right) \rightarrow 2 \hbar \Omega_x ^{(1)} \, (\, \hat{\sigma}_x^{ab} - \hat{\sigma}_x^{cd}\,) \, + \, 2 \hbar \Omega_y ^{(1)} \,(\,\hat{\sigma}_y^{ab} - \hat{\sigma}_y^{cd} \,)\, + \, 2 \hbar \Omega_z ^{(1)}\, (\, \hat{\sigma}_z^{ab} - \hat{\sigma}_z^{cd} \,),
\end{equation}
\begin{equation}
\label{eqsA06B}
i \hat{\beta} \hat{\bm{\alpha}} \cdot \left( \mu \, \frac{\bm{\mathcal{E}}}{c} \right) \rightarrow 2 \hbar \Omega_x ^{(2)}\, (\, -\hat{\sigma}_y^{ad} - \hat{\sigma}_y^{bc}\,) \, + \, 2 \hbar \Omega_y ^{(2)}\,(\, \hat{\sigma}_x^{bc} - \hat{\sigma}_x^{ad}\,) \, + \,2 \hbar \Omega_z ^{(2)} \,(\, \hat{\sigma}_y^{bd} - \hat{\sigma}_y^{ac} \,).
\end{equation}
\end{subequations}
To use the maps from Eqs.~(\ref{eqsA04}), (\ref{eqsA05}) and (\ref{eqsA06}) to reproduce the Hamiltonian (\ref{eqsA00}), the relations between the observable and Dirac-like parameters are established by
\begin{eqnarray}
\frac{\mu\, \mathcal{E}_j}{c} &=& 2 \hbar \Omega_j ^{(2)}, \hspace{1.3 cm} \kappa\, \mathcal{E}_j = 2 \hbar \Omega_j ^{(1)}, \nonumber \\
c &=& 2 \eta \Delta \tilde{\Omega}, \hspace{1.3 cm} m c^2 = 2 \hbar \delta,
\end{eqnarray}
through which one identifies a one-to-one correspondence between the Hamiltonian from (\ref{eqsA00}) and the sum of the interactions (\ref{eqsA04}), (\ref{eqsA05}) and (\ref{eqsA06}). The Dirac equation with tensor and pseudotensor potentials is thus simulated by the four internal levels of the trapped ion, and the eigenstates of (\ref{eqsA00}), $\vert \psi_{n,s} \rangle$ ($n,\,s = 0,\,1$) are therefore encoded by the superposition of the internal ionic states,
\begin{equation}
\label{eqsA07}
\vert \psi_{n,s} \rangle \rightarrow M^a_{n,s} \vert a \rangle + M^b_{n,s} \vert b \rangle + M^c_{n,s} \vert c \rangle + M^d_{n,s} \vert d \rangle.
\end{equation}

Alkali ions such as Mg$^+$, Ca$^+$ and Sr$^+$ exhibit hyperfine levels that can be used as a platform for the above described setup.
Fig.~\ref{eqsfig:level} also pictorially illustrates typical hyperfine levels of the $2 s^2 \, S_{1/2}$ ground state of such typical alkali ions.
Considering that two intrinsic degrees of freedom are strictly related to the projection of the total angular momentum ${\bm F}$ onto the trapping magnetic field ${\bm M}$, one may adopt a two-{\em qubit} assignment to the internal ionic states as
\begin{eqnarray}
\label{eqsA08}
\vert a \rangle &\equiv& \vert 0 \, 0 \rangle, \hspace{1.5 cm} \vert b \rangle \equiv \vert 0 \, 1 \rangle, \nonumber \\
\vert c \rangle &\equiv& \vert 1 \, 0 \rangle, \hspace{1.5 cm} \vert d \rangle \equiv \vert 1 \, 1 \rangle,
\end{eqnarray}
which shall be used in the following calculations.

\section{Eigenstates of the Dirac Hamiltonian}

The systematic engineering of entangled structures involving the internal ionic levels (described through the correspondence between Dirac-like and the trapped ion systems introduced in the previous section) demands for a deeper analysis of a larger class of bi-spinor interactions \cite{n009}.
For some classes of Poincar\'e invariant Dirac-like interactions, the Hamiltonian eigenstates may exhibit a naturally entangled structure which can be directly computed from stationary pure states.
That is not the case of Hamiltonian systems driven by an electromagnetic field minimal coupling (via ${\bm p} \to {\bm p} - {\bm A}$).

In fact, the invariance of the Dirac equation under Poincar\'e transformations restricts the inclusion of additional external fields to the Dirac Hamiltonian by scalar, pseudoscalar, vector, pseudovector, tensor and pseudotensor potentials, once they are typified by their transformation properties \cite{n015}.
The Hamiltonian (\ref{eqsA00}) includes both tensor and pseudotensor potentials that describe the non-minimal coupling with an external constant electric field.
In particular, it also exhibits algebraic properties which allows for obtaining pure states as Hamiltonian eigenstates \cite{n009}.
By assuming from this point that one has set $c = \hbar = 1$ for simplifying reasons, from Eq.~(\ref{eqsA00}), one has
\begin{equation}
\hat{\mathcal{H}}^2 = g_1\, \hat{I}_4 + 2 \hat{\mathcal{O}},
\end{equation}
where $\hat{\mathcal{O}}$ is a traceless operator given by
\begin{equation}
\hat{\mathcal{O}} = m\,\kappa\, \hat{\bm{\Sigma}} \cdot \bm{\mathcal{E}} + \mu \, \hat{\beta}\,\hat{\bm{\Sigma}} \cdot \, (\, \bm{p} \times \bm{\mathcal{E}} \,) - i \kappa\,\hat{\beta} \,\hat{\bm{\alpha}} \cdot (\, \bm{p} \times \bm{\mathcal{E}} \,),
\end{equation}
with
\begin{equation}
\hat{\mathcal{O}}^2 = \frac{1}{4}(\hat{\mathcal{H}}^2 - g_1\, \hat{I}_4)^2 = g_2\, \hat{I}_4,
\end{equation}
and
\begin{eqnarray}
g_1 &=& \frac{1}{4}\Tr[\hat{\mathcal{H}}^{2}] = p^2 + m^2 + (\kappa^2 + \mu^2) \mathcal{E}^2, \nonumber \\
g_2 &=& \frac{1}{16}\Tr\left[\left(\hat{\mathcal{H}}^2- \frac{1}{4}\Tr[\hat{\mathcal{H}}^{2}]\right)^{2}\right] = m^2 \kappa^2 \mathcal{E}^2 + (\mu^2 + \kappa^2)(\bm{p}\times \bm{\mathcal{E}})^2.
\end{eqnarray}
By using a simple \textit{ansatz} (c. f. Ref.~\cite{n009}), one constructs the corresponding Hamiltonian eigenvalues through the density operators,
\begin{equation}
\label{eqsB01}
\varrho_{n,s} = \frac{1}{4} \left( \, \hat{I}_4 + \frac{(-1)^s}{\sqrt{g_2}} \hat{\mathcal{O}} \, \right) \left(\, \hat{I}_4 + \frac{(-1)^n}{\vert \, \lambda_{n,s} \, \vert} \hat{\mathcal{H}} \, \right),
\end{equation}
which indeed correspond to pure state solutions of the stationary Liouville equation $[\varrho_{n,s}, \hat{\mathcal{H}}] = 0$.
Once the states from (\ref{eqsB01}) are identified with the pure states that provide solutions for the Dirac-like equation, i. e. their eigenspinor solutions, the eigenvalue parameter, $\lambda$, can be evaluated by
\begin{equation}
\lambda_{n,s} = (-1)^n \sqrt{ g_1 + 2 \,(-1)^s \sqrt{g_2}},
\end{equation}
which therefore corresponds to the averaged energy $E_{n,s}=\mbox{Tr}[\hat{\mathcal{H}} \,\varrho_{n,s}] =\lambda_{n,s}$.

By identifying the state given by (\ref{eqsB01}) as a composite quantum system, one can compute entanglement and additional quantum correlations between the pertinent subsystems.
For Dirac equation solutions, these quantum correlations are related to \textit{spin polarization} and {\em intrinsic parity} internal degrees of freedom, as they have been identified in the context of the above mentioned Poicar\'e invariant external couplings \cite{n009}.
Therefore, the \textit{spin-parity} entanglement reflects the $\mbox{SU}(2)\otimes \mbox{SU}(2)$ bi-spinor structure of these solutions (c. f. the Appendix).

The representation (\ref{eqsB02}) from the Appendix supports the interface between relativistic quantum mechanics and quantum information theory \cite{n018}, where the discrete degrees of freedom are associated to a system $\mathcal{S}$ composed by two subsystems, $\mathcal{S}_1$ (\textit{spin} system) and $\mathcal{S}_2$ (\textit{intrinsic parity} system), embedded into a composite Hilbert space $\mathbb{H} = \mathbb{H}_1 \otimes \mathbb{H}_2$ with $\mbox{dim}\, \mathbb{H}_1 = \mbox{dim}\, \mathbb{H}_2 = 2$.
The corresponding bi-partite states are indeed two-{\em qubit} states, for which, when external fields are included into the Dirac dynamics, the correlation content of bi-spinors changes.
Through the \textit{ansatz} from (\ref{eqsB01}), the entire entanglement/correlation content can be obtained.

Quantum entanglement can be read as consequence of the superposition principle, and it is related to the concept of separability \cite{n019}. A bi-partite state described by a density operator $\rho$ is said separable if \cite{n019}
\begin{equation}
\rho = \sum_{i} w_i \, \sigma_i ^{(1)} \otimes \tau_i ^{(2)},
\end{equation}
where $\sigma_i ^{(1)} \in \mathbb{H}_1$, $\tau_i ^{(2)} \in \mathbb{H}_2$ and $\sum_i w_i = 1$. If a state is not separable, then it is entangled. For pure states, the quantum entanglement can be quantified by the entanglement entropy $E_{VN}[\rho]$ computed through the von Neumann entropy of a subsystem\cite{n020},
\begin{equation}
E_{vN}[\rho] = S[\rho_2] = - \mbox{Tr}_2[\rho_2 \log_2 \rho_2] = S[\rho_1] = - \Tr_1[\rho_1 \log_2 \rho_1],
\end{equation}
where the above equality is guaranteed by the Schmidt decomposition theorem, which asserts that, for pure states $\rho$, the reduced density operators, $\rho_{1 \, (2)} = \Tr_{2 (1)}[\rho]$, have identical eigenvalues and, if the state is entangled, then either $\rho_{1(2)}$ are mixed states \cite{n020}. Other entanglement quantifier often considered is the quantum concurrence, $C[\rho]$, whose definition is primarily related to the calculation of entanglement of formation of two-{\em qubit} mixed states \cite{n021}.
For pure states, concurrence has a simplified formula once it has been demonstrated that any two-{\em qubit} system can be written in the form of
\begin{equation}
\rho = \frac{1}{4} \left[ I_4 + (\hat{\bm{\sigma}}^{(1)} \otimes \hat{I}_2^{(2)}) \cdot \bm{a}_1 + (\hat{I}_2^{(1)} \otimes \hat{\bm{\sigma}}^{(2)}) \cdot \bm{a}_2 + \displaystyle \sum_{i,j = 1}^3 t_{ij} (\hat{\sigma}_i^{(1)} \otimes \hat{\sigma}_j^{(2)}) \right],
\end{equation}
where $\hat{\sigma}_i$ are the Pauli matrices, $[T]_{ij} = t_{ij}$ is the correlation matrix and $\bm{a}_{1 \, (2)}$ are the Bloch vectors of the corresponding subsystem. For pure states, $a_1 ^2 = a_2 ^2$ and the concurrence is given in terms of the Bloch vectors by
\begin{equation}
\label{eqsB03}
\mathcal{C}[\varrho] = \sqrt{1 - a_{1}^2} = \sqrt{1 - a_{2}^2}.
\end{equation}
For the correspondence established by (\ref{eqsA08}), the Bloch vector $\bm{a}_2 \def \bm{a}$ for the state (\ref{eqsB01}) is given by
\begin{equation}
\label{eqsB0001}
\bm{a}_2 = \mbox{Tr}_1 [\hat{\bm{\Sigma}} \varrho_{n,s}]= \frac{(-1)^s \, m}{\sqrt{g_2}} \, \left[\kappa\, \bm{\mathcal{E}} + \frac{(-1)^n \, \mu \, (\, \bm{p}\times \bm{\mathcal{E}} \,) }{\vert \, \lambda_{n, s} \, \vert} \, \right],
\end{equation}
from which the concurrence is evaluated through (\ref{eqsB03}).

Finally, a suitable correspondence between the critical point values of the averaged chirality and \textit{spin-parity} entanglement can be identified \cite{n011}.
The chirality of an state $\varrho_{n,s}$ is computed through the average value of the operator $\hat{\gamma}_5 = i \hat{\gamma}_0 \hat{\gamma}_1 \hat{\gamma}_2 \hat{\gamma}_3 = - i \hat{\alpha}_x \, \hat{\alpha}_y \, \hat{\alpha}_z \equiv \hat{\sigma}^{(1)}_x \otimes \hat{I}_2^{(2)} $,
\begin{equation}
\langle \, \hat{\gamma}_5 \, \rangle = \mbox{Tr}[\, \hat{\gamma}_5 \varrho_{n,s}\,].
\end{equation}

By following the definitions from Eq.~(\ref{eqsAAA}), and the correspondence from Eqs.~(\ref{eqsA04})-(\ref{eqsBBB}), in terms of the ionic states, one obtains $\hat{\gamma}_5 = \vert a \rangle \langle d \vert + \vert d \rangle \langle a \vert + \vert b \rangle \langle c \vert + \vert c \rangle \langle b \vert$, such that the averaged chirality for a pure state can be related to transition probabilities.
Since the probability of measuring a pure state $\vert \psi \rangle$ in the maximal superposition $(\vert a \rangle + \vert d \rangle)/\sqrt{2}$ is given by
\begin{equation}
P_{ad} = \left \vert \left(\,\frac{\langle a \vert + \langle d \vert}{\sqrt{2}} \, \right) \, \vert \psi \rangle \, \right \vert^2,
\end{equation}
(with an analogous definition for $P_{cb}$), after simple math manipulations, and using the fact that $\displaystyle \sum_{i = a, \, ... \,\, d} \vert \langle i \vert \psi \rangle \vert^2 = 1$, one obtains the following relation
\begin{equation}
\label{eqsB06}
\langle \, \hat{\gamma}_5 \, \rangle = 2 (P_{ad} + P_{cb}) - 1.
\end{equation}
It means that the averaged chirality is related to the probabilities of measuring the system in maximal superpositions of $\{\vert a \rangle,\,\vert d \rangle\}$ and $\{\vert b \rangle,\,\vert c \rangle\}$.
In particular, if the quantum state superposition results into $\langle \, \hat{\gamma}_5 \, \rangle = -1$, one has $P_{ad} = P_{cb}=0$ and one should have a quantum superposition between cat-like states, $(\vert a \rangle - \vert d \rangle)/\sqrt{2}$, and Werner-like states, $(\vert c \rangle - \vert b \rangle)/\sqrt{2}$ (c. f. Eq.~\eqref{eqsA08}).

\begin{figure}
\centering
\includegraphics[width = 8 cm]{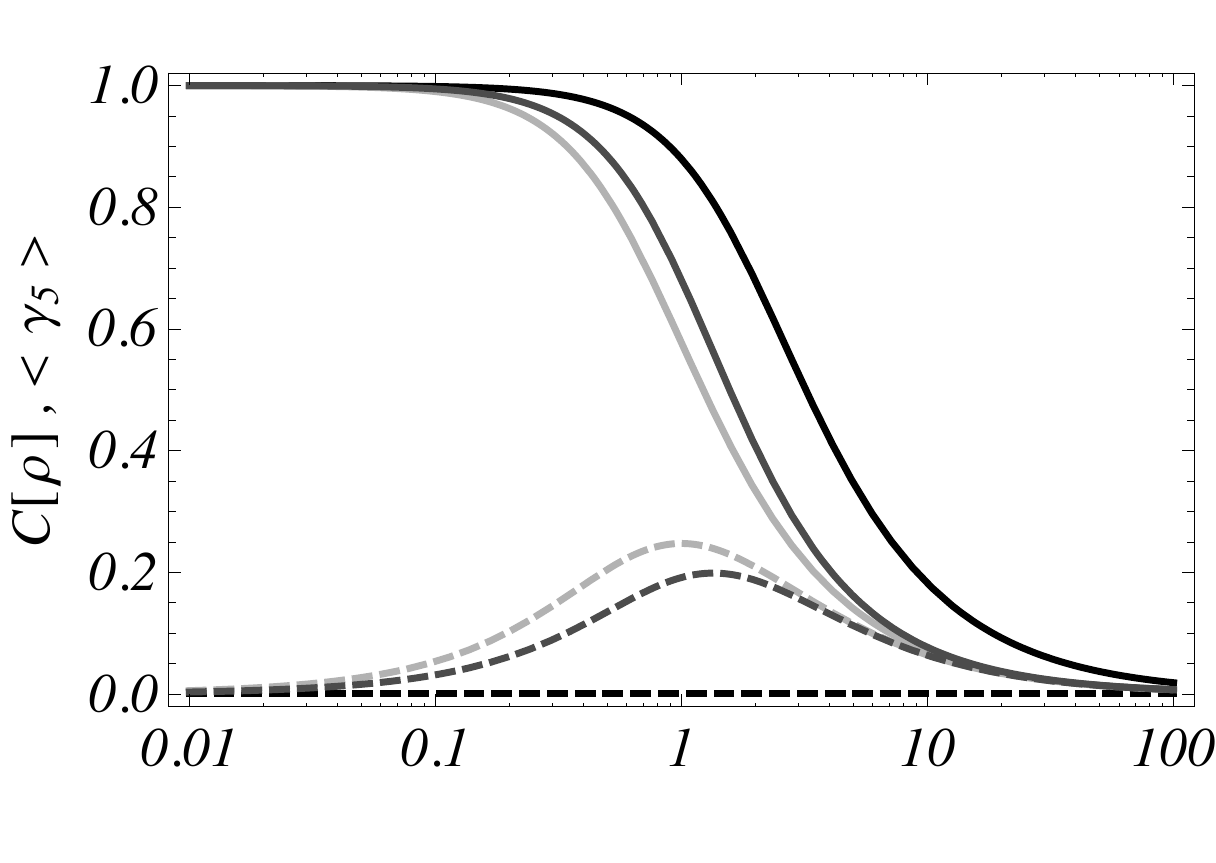}
\includegraphics[width = 8.3 cm]{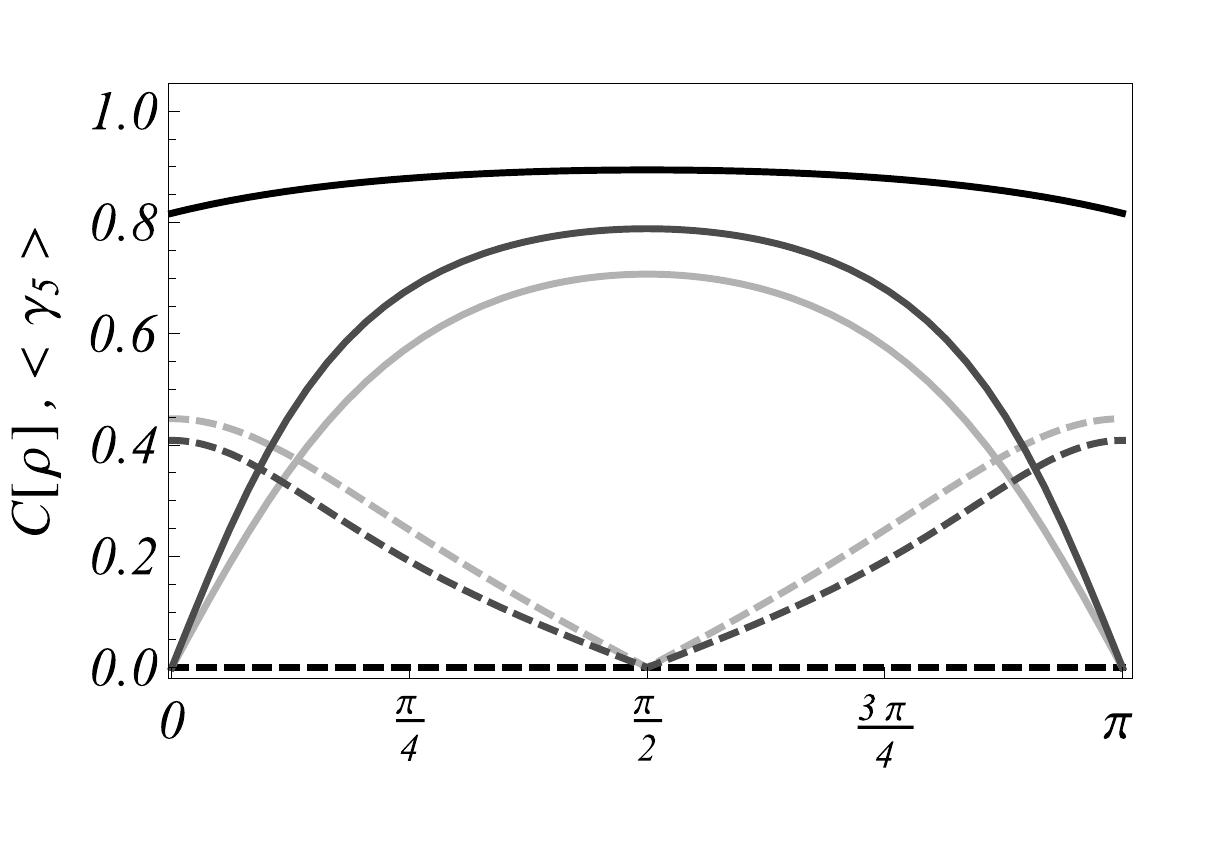}
\includegraphics[width = 8 cm]{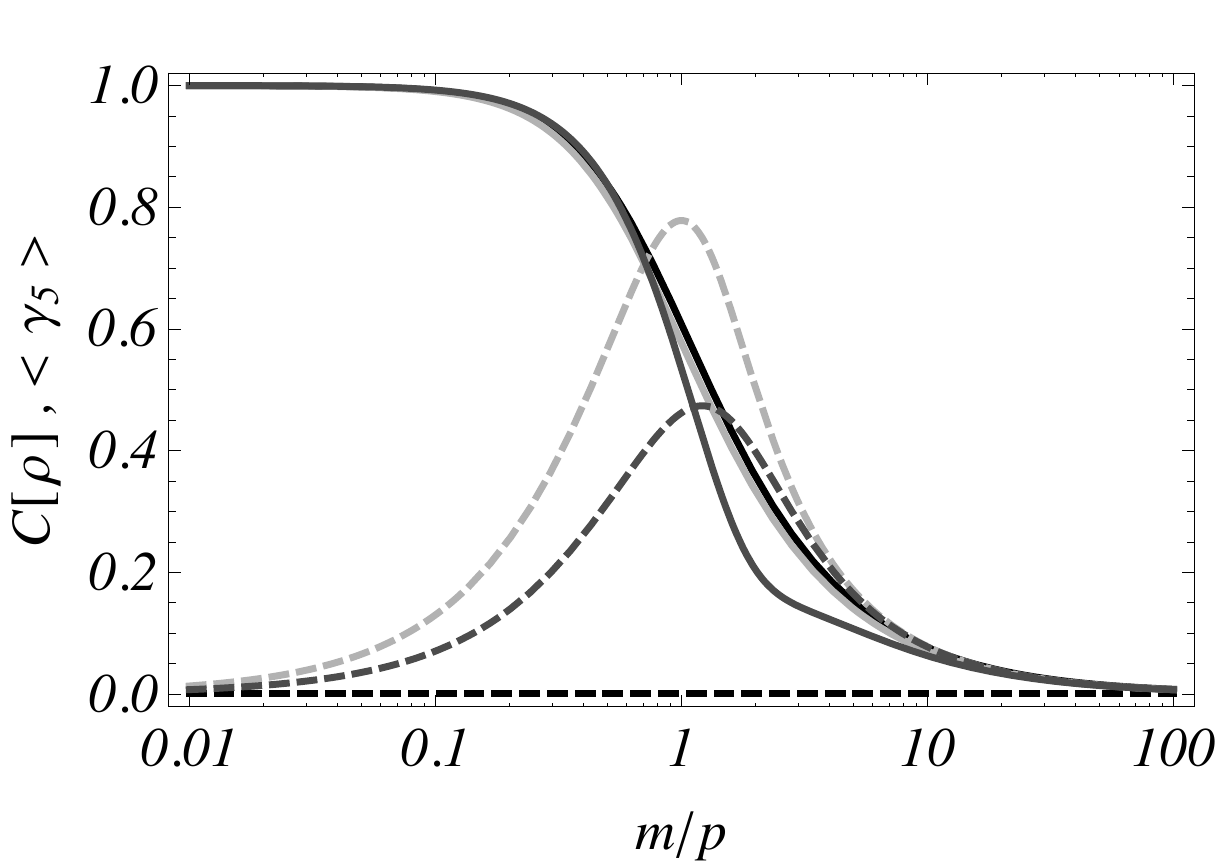}
\includegraphics[width = 8.3 cm]{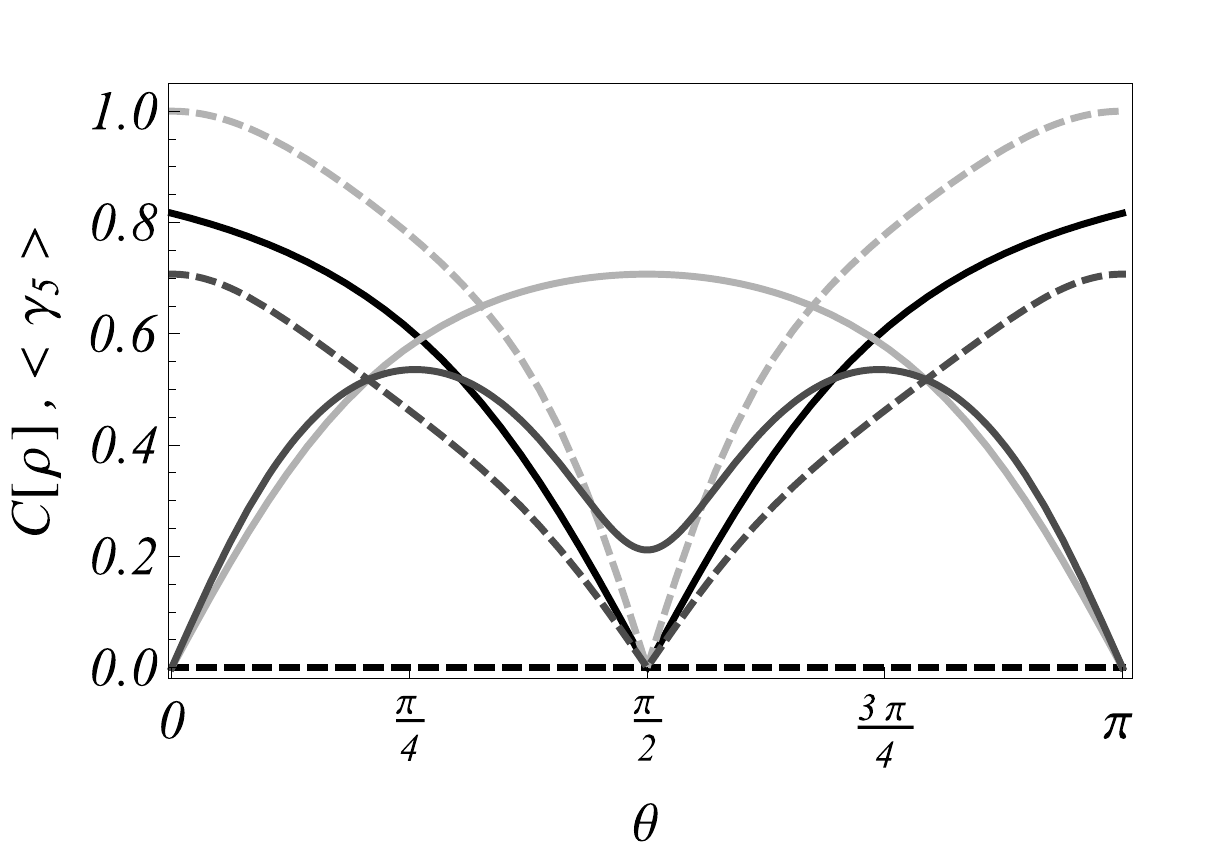}
\caption{Quantum concurrence, $C[\rho]$ (solid lines), and modulus of the averaged chirality,$\vert\langle\hat{\gamma}_5\rangle\vert$ (dashed lines), for the density matrix given by Eq.~(\ref{eqsB01}) (for one-dimensional propagation according to (\ref{eqsB05})) as functions of $m/p$ (in natural units $\sim m(c)/p$) for $\theta = \pi/4$ (left column), and of $\theta$ for $m/p = 1$ (right column). The first and second rows corresponds to $s=0$ and $1$, respectively. The plots are for $(\kappa,\, \mu) = $ $( 0 , 1 )$ (black), $(1, 0 )$ (dark gray), and $(1, 1)$ (light gray).
Notice that the entanglement is a strictly decreasing function of $m/p$ that vanishes for $m/p \rightarrow \infty$ (non-relativistic limit).
On the other hand, if $p \gg m$ the state is maximally entangled. For $\kappa = 0$, the state averaged chirality vanishes and, in the converse case, the maximum point of $\vert \langle \hat{\gamma}_5 \rangle \vert$ corresponds to an inflection point for the concurrence. As function of $\theta$, the averaged chirality vanishes for $\theta = \pi/2$, which also corresponds to a local critical point for the concurrence. For instance, the state is separable for $\kappa =1$, $\mu = 1$ and $\theta = \pi/2$.}
\label{eqsfig:01}
\end{figure}

Let one extends such analysis to the particular configuration of an one-dimensional propagation along the $x$ axis, with the electric field lying in the plane $x-y$, for which
\begin{eqnarray}
\label{eqsB05}
\bm{\mathcal{E}} &=& \mathcal{E}(\, \cos{\theta}\,\bm{i} + \sin{\theta} \,\bm{j} \,),
\end{eqnarray}
with $\bm{p} = p \,\bm{i}$, in the scenario where $\bm{p} \times \bm{\mathcal{E}} = p \, \mathcal{E} \sin{\theta} \, \bm{k}$ and $\bm{i},\bm{j},\bm{k}$ are unitary vectors.
In this case, simplified expressions for $g_2$, $\lambda_{n,s}$, and for the modulus of the Bloch vector are given by
\begin{subequations}
\begin{eqnarray}
g_2 &=& \mathcal{E}^2 [\, m^2 \kappa^2 + (\mu^2 + \kappa^2)\,p^2 \sin^2{\theta} \,], \\
\lambda_{n,s} &=& (-1)^n\bigg[\, p^2 + m^2 + (\kappa^2 + \mu^2)\mathcal{E}^2 + 2 (-1)^s \mathcal{E}\sqrt{m^2 \kappa^2 + (\mu^2 + \kappa^2)\,p^2 \sin^2{\theta}}\, \bigg]^ {1/2}, \\
a_2 ^2 &=& \frac{m^2 }{m^2 \kappa^2 + (\mu^2 + \kappa^2)\,p^2 \sin^2{\theta} } \, \left[\, \kappa^2 + \frac{ \mu^2 \, p^2 \, \sin^2{\theta} }{ \lambda^2_{n, s}} \, \right],
\end{eqnarray}
\end{subequations}
and the averaged chirality is given by
\begin{equation}
\label{eqsB04}
\langle \hat{\gamma}_5 \rangle = \frac{(-1)^{n + s} m \, p \, \kappa\, \cos{\theta} }{\vert \, \lambda_{n,s} \, \vert \, \sqrt{ m^2 \kappa^2 + (\mu^2 + \kappa^2)\,p^2 \sin^2{\theta}} }.
\end{equation}
The absolute value of (\ref{eqsB04}) and the concurrence are depicted in Fig.~\ref{eqsfig:01} as functions of $m/p$ (left column) and $\theta$ (right column) for $(\kappa,\,\mu) = $ $(0,1)$, $(1, 0)$ and $(1,1)$. Concurrence is a strictly decreasing function of $m/p$ and the state is separable in the non-relativistic limit, i. e. for $m/p \rightarrow \infty$. On the other hand, for $m \ll p$, i. e. in the ultra-relativistic regime, the state is maximally entangled. The averaged chirality has a maximum that corresponds to an inflection point of the concurrence.
In the absence of the pseudotensor interaction, i. e. for $\kappa = 0$, $\langle \hat{\gamma}_5 \rangle$ vanishes. The concurrence has a local extreme value for $\theta = \pi/2$, such that for $\kappa = 0$ the state is separable. For this value of $\theta$, the chirality always vanishes as can be directly inferred from (\ref{eqsB04}).

\section{Recovering the internal ionic state dynamics}

Once entangling and chiral properties of the states $\varrho_{n,s}$ have been assigned, the straightforward connection to the dynamics of the internal ionic states can now be obtained.
By following a step-by-step construction, the coefficients of the quantum superposition from (\ref{eqsA07}), $M^{i} _{n,s}$ ($i = a,\, b,\, c,\, d$), compose a matrix $M$ that connects the Dirac bi-spinor basis, $\{ \vert \, \psi_{n,s} \, \rangle \}$ ($n,s = $ $0,1$), to the ionic state basis, $\{ \vert\, i \,\rangle \}$.
The expressions for $\vert M^i_{n,s} \vert$ can be obtained by the diagonal elements of the density operator $\varrho_{n,s}$,
\begin{equation}
\vert M^i_{n,s} \vert = \sqrt{\mbox{Tr}[\varrho_{n,s} \vert i \rangle \langle i \vert] }.
\end{equation}
In the same fashion, the relative phases between $\vert i \rangle$ and $\vert j \rangle$, $e^{i \Delta \phi^{ij}_{n,s}}$, are extracted from the off diagonal elements of the density operator,
\begin{equation}
e^{i \Delta \phi^{ij}_{n,s}} = \frac{\mbox{Tr}[\varrho_{n,s} \vert i \rangle \langle j \vert]}{ \vert\, M^{i}_{n,s} \, \vert \, \vert \, M^{j}_{n,s} \, \vert}.
\end{equation}
Apart from a global phase factor which has been assumed to be $e^{i \phi_a}$, one thus determines the Hamiltonian eigenstates, $\vert \psi_{n,s} \rangle$, as given by
\begin{equation}
\vert \psi_{n,s} \rangle = e^{i \phi_a}\bigg{[} \vert M^a_{n,s} \vert\, \vert a \rangle + \vert M^b_{n,s} \vert\, e^{- i \Delta \phi^{ab}_{n,s}} \, \vert b \rangle + \vert M^c_{n,s}\vert \, e^{- i \Delta \phi^{ac}_{n,s}} \, \vert c \rangle + \vert \, M^d_{n,s} \, \vert e^{- i \Delta \phi^{ad}_{n,s}} \vert d \rangle \bigg{]}.
\end{equation}
By introducing the constraints from Eqs.~(\ref{eqsB05})-(\ref{eqsB04}), the expressions for the superposition coefficients can be written as
\begin{subequations}
\begin{equation}
\vert M^a_{n,s} \vert = \frac{1}{2} \left[\, 1 + \frac{(-1)^n m }{\vert \lambda_{n,s} \vert} + \frac{(-1)^s p \mu\, \mathcal{E} \sin{\theta}}{\sqrt{g_2}} + \frac{(-1)^{n+s} m (\, p \, \mu \,\mathcal{E} \sin{\theta} + \kappa^2 \mathcal{E}^2 \,)}{\sqrt{g_2} \, \vert \lambda_{n,s} \vert} \, \right]^{1/2},
\end{equation}
\begin{equation}
\vert M^b_{n,s} \vert = \frac{1}{2} \left[\, 1 + \frac{(-1)^n m }{\vert \lambda_{n,s} \vert} - \frac{(-1)^s p \mu\, \mathcal{E} \sin{\theta}}{\sqrt{g_2}} + \frac{(-1)^{n+s} m (\, p \, \mu \,\mathcal{E} \sin{\theta} + \kappa^2 \mathcal{E}^2 \,)}{\sqrt{g_2} \, \vert \lambda_{n,s} \vert} \, \right]^{1/2},
\end{equation}
\begin{equation}
\vert M^c_{n,s} \vert = \frac{1}{2} \left[\, 1 - \frac{(-1)^n m }{\vert \lambda_{n,s} \vert} - \frac{(-1)^s p \mu\, \mathcal{E} \sin{\theta}}{\sqrt{g_2}} + \frac{(-1)^{n+s} m (\, p\,\mu \,\mathcal{E} \sin{\theta} + \kappa^2 \mathcal{E}^2 \,)}{\sqrt{g_2}\,\vert \lambda_{n,s} \vert} \, \right]^{1/2},
\end{equation}
\begin{equation}
\vert M^d_{n,s} \vert = \frac{1}{2} \left[\, 1 - \frac{(-1)^n m }{\vert \lambda_{n,s} \vert} + \frac{(-1)^s p \mu\, \mathcal{E} \sin{\theta}}{\sqrt{g_2}} + \frac{(-1)^{n+s} m (\, p \, \mu\, \mathcal{E} \sin{\theta} + \kappa^2 \mathcal{E}^2\, )}{\sqrt{g_2}\, \vert \lambda_{n,s} \vert} \, \right]^{1/2},
\end{equation}
\end{subequations}
and the corresponding relative phases are thus given by
\begin{subequations}
\begin{eqnarray}
\label{eqsC01A}
e^{-i \Delta \phi^{ab}_{n,s}} &=& \frac{\kappa\, \mathcal{E}}{4 \vert M^a_{n,s} \vert \, \, \vert M^b_{n,s} \vert } \bigg[\, \frac{(-1)^n e^{i \theta}}{\vert \, \lambda _{n,s} \, \vert} +\frac{(-1)^s m e^{i \theta}}{\sqrt{g_2}} \nonumber \\
 &&\qquad\qquad+ \frac{(-1)^{n+s} \, \,(p\, \sin{\theta} (\mu \, \mathcal{E} e^{i \theta} +i p ) + e^{i \theta} (\, m^2- p \,\mu\, \mathcal{E} \sin{\theta} \,)\,)}{\sqrt{g_2} \, \vert \, \lambda_{n,s} \, \vert} \, \bigg],\\
\label{eqsC01B}
e^{-i \Delta \phi^{ac}_{n,s}} &=& \frac{ i \, \kappa\, \mathcal{E} }{4\, \sqrt{g_2} \, \vert M^a_{n,s} \vert \, \, \vert M^c_{n,s} \vert} \bigg[ (-1)^s p\, \sin{\theta}\nonumber \\
&&\qquad\qquad+ \frac{(-1)^{n+s} \, m( p \, \sin{\theta} - e^{- i \theta} ( \mu\, \mathcal{E} e^{i \theta} + i p))}{ \, \vert \, \lambda_{n,s} \, \vert} \bigg],\\
e^{- i \Delta \phi^{ad}_{n,s}} &=& - \frac{i}{4 \, \vert \, \lambda_{n,s} \, \vert \, \vert M^a_{n,s} \vert \, \, \vert M^d_{n,s} \vert } \bigg[(-1)^n (\mu\, \mathcal{E} e^{i \theta} + i p) \nonumber \\
&&\qquad\qquad+ \frac{(-1)^{n+s} \mathcal{E}(p \, \kappa^2 \mathcal{E} \sin{\theta} e^{i \theta} + p \,\mu \, \sin{\theta} (\mu\, \mathcal{E} e^{i \theta} + i p))}{\sqrt{g_2} \, \vert \, \lambda_{n,s} \, \vert} \bigg].
\end{eqnarray}
\end{subequations}

As expected, from the above expressions, after such exhaustive math manipulations, one verifies that $$\displaystyle \sum_{i = a,\, ... \, , d} \vert M^i_{n,s} \vert^2 = 1,$$
which ratifies that the states $\vert \psi_{n,s} \rangle$ are all normalized.
Of course, the spinor states are eigenstates of $\hat{\mathcal{H}}$, therefore their temporal evolution reads (for $\vert \psi_{n,s} (t = 0) \rangle \equiv \vert \psi_{n,s} \rangle$)
\begin{equation}
\vert \psi_{n,s} (t) \rangle = e^{- i \hat{\mathcal{H}} t} \vert \psi_{n,s} \rangle = e^{- i \lambda_{n,s} t} \vert \psi_{n,s} \rangle.
\end{equation}

Analogously, in order to describe the dynamics of an internal ionic level, $\vert j \rangle$, one may write it as a superposition of bi-spinor states $\vert \psi_{n,s} \rangle$ as
\begin{equation}
\vert j \rangle = \displaystyle \sum_{(n,s) = 0,1} W^j _{n,s} \vert \psi_{n,s} \rangle,
\end{equation}
where the elements $W^j _{n,s}$ compose the inverse matrix {\bf W} $=$ {\bf M}$^{-1}$ such that $$\displaystyle \sum_{(n,s) = 0,1} \vert W^i_{n,s} \vert^2 = 1.$$
Notice that the ionic states are not Hamiltonian eigenstates, and their temporal evolution are given by
\begin{equation}
\label{eqsC02}
\vert \, j (t) \rangle = e^{-i \hat{\mathcal{H}}} \, \vert \, j \rangle = \displaystyle \sum_{(n,s) = 0,1} W^{j}_{n,s} e^{- i \lambda_{n,s} t} \vert \psi_{n,s} \rangle,
\end{equation}
which, for $\vert j (t=0) \rangle \equiv \vert j \rangle$, gives a typical pattern of the quantum oscillation phenomena for a four level system.
A state initially prepared as $\vert j \rangle$ oscillates and can be converted into other states, $\vert k \rangle \neq \vert j \rangle$.
By defining the projector $\hat{P}_k = \vert k \rangle \langle k \vert$ onto a generic $\vert k \rangle$ ionic state, the probability of measuring the trapped ion system in such a configurational state is given by
\begin{eqnarray}
P_{j \rightarrow k} (t) = \mbox{Tr}[\vert j (t) \rangle \langle j(t) \vert \, P_k] = \displaystyle \sum_{(n,s) = 0,1} \sum_{(m,l) = 0,1} W^j _{n,s} \, W^k _{m,l} \big{(}\, W^j _{m,l} \, \big{)}^* \big{(}\, W^k _{n,s} \, \big{)}^* \, e^{- i (\lambda_{n,s} - \lambda_{m,l}) t}.
\end{eqnarray}

\begin{figure}
\includegraphics[width =13.7 cm]{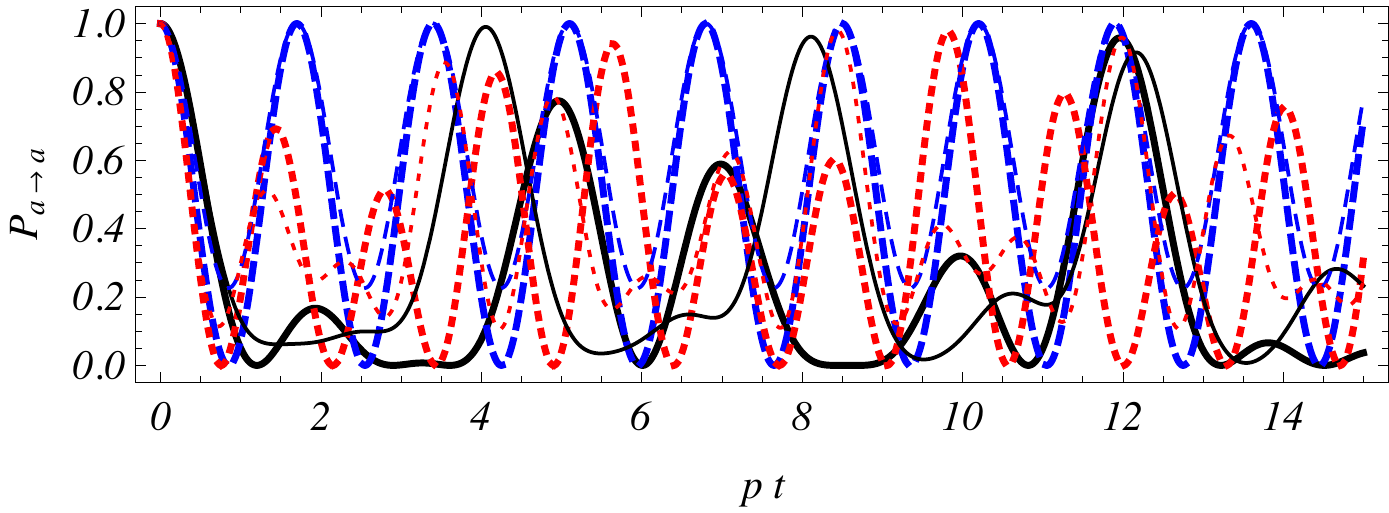}
\vspace{- .5cm}
\includegraphics[width =13.7 cm]{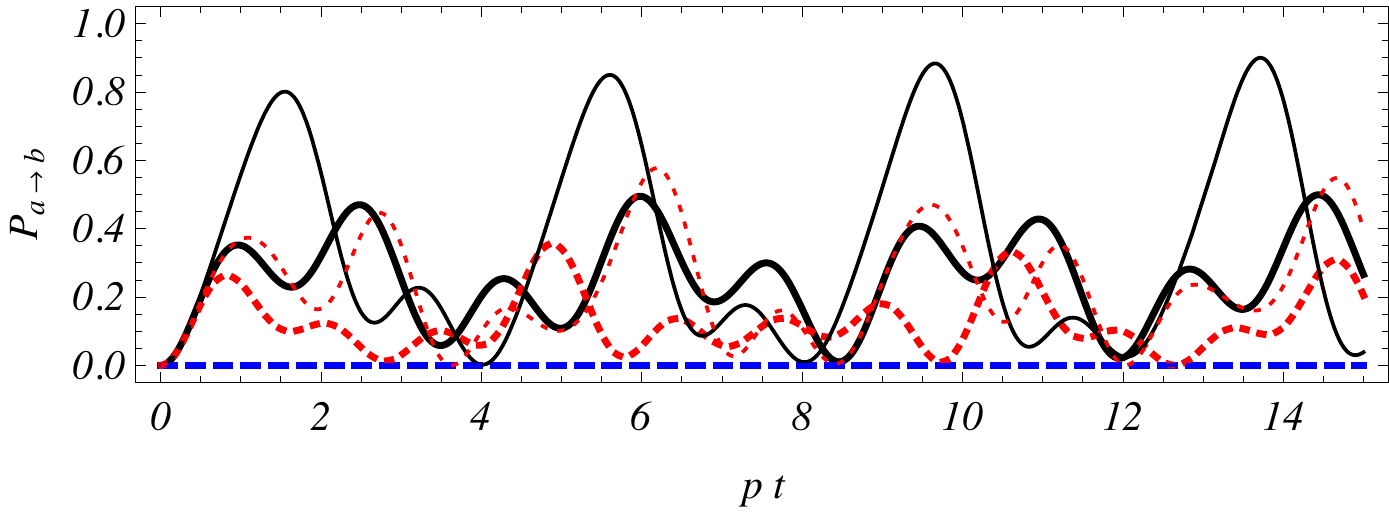}
\vspace{- .5cm}
\includegraphics[width =13.7 cm]{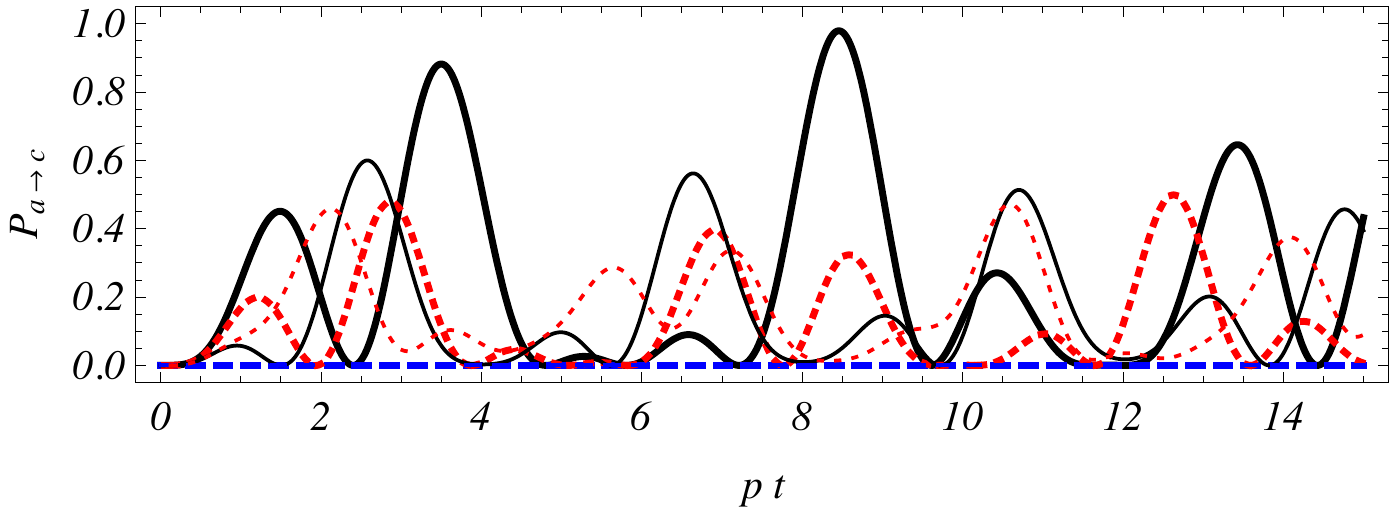}
\vspace{- .5cm}
\includegraphics[width =13.7 cm]{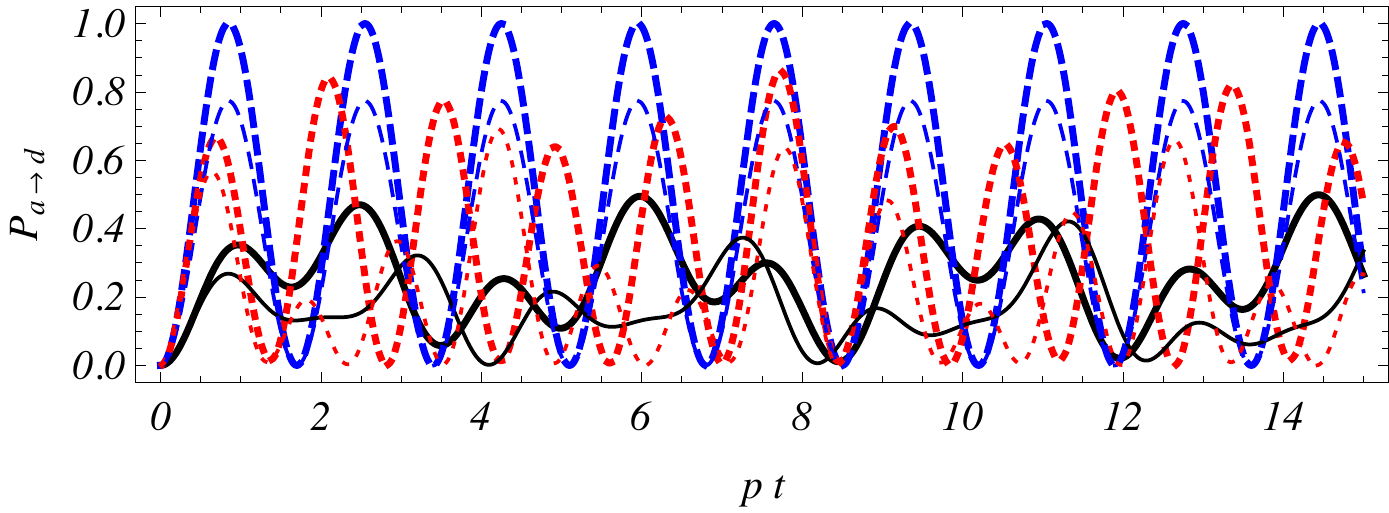}
\caption{Transition probabilities, $P_{a\to a,b,c,d}$ as functions of a dimensionless parameter $p\,t$ (in natural units $\sim p\,t (c/\hbar)$.
Thick lines are for $m=0$, and thin lines are for $m=1$. The plots are for $(\kappa ,\, \mu)= (1,0)$ (solid lines), $(0,1)$ (dashed lines) and $(1,1)$ (dotted lines).
Since the transition probabilities depend on a combination of harmonic functions with different frequencies, they do not generally exhibit an identifiable periodicity.
One also notice that for $\kappa = 0$ (dashed lines), the probabilities $P_{a \rightarrow b}$ and $P_{a \rightarrow c}$ are null (as it follows from Eqs.~(\ref{eqsC01A})-(\ref{eqsC01B})), and only $\vert a \rangle$ and $\vert d \rangle$ are relevant for the dynamics.}
\label{eqsfig:02}
\end{figure}

\begin{figure}
\includegraphics[width = 15 cm]{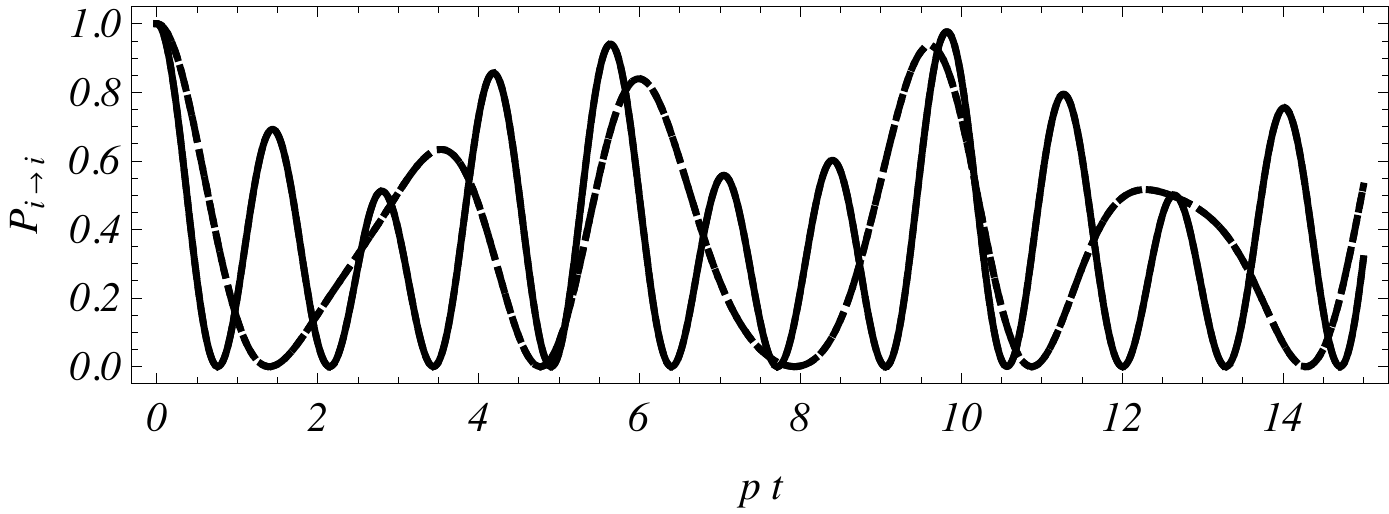}
\vspace{- .5cm}
\includegraphics[width = 15 cm]{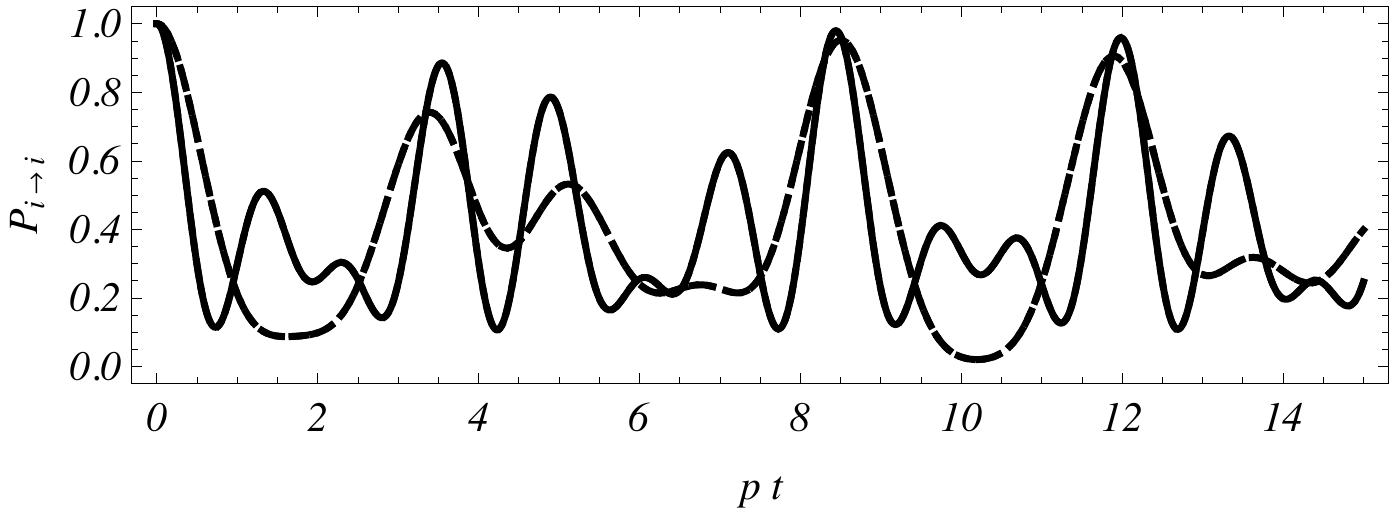}
\caption{Survivor probabilities, $P_{i \rightarrow i}$, for initial states $\vert i \rangle = \vert a \rangle$ (solid lines) and $\vert i \rangle = \vert d \rangle$ (dash-dotted lines), as functions of $p\,t$. The plots are for $m = 0$ (first plot), which corresponds to the suppression of the detuning effect due to $\delta$, and for $m=1$ (second plot), with $\kappa = \mu = 1$.}
\label{eqsfig:03}
\end{figure}

Fig.~\ref{eqsfig:02} depicts the survivor probability $P_{a \rightarrow a}$ and the transition probabilities $P_{a \rightarrow b,c,d}$ for the state $\vert a(t) \rangle$, as function of a dimensionless parameter, $p\,t$ (which in natural units is $p\,t (c/\hbar)$), for $\theta = \pi/4$. The choice of $\theta$ is arbitrary and it does not affect qualitatively the results.
Since the ratio between the self-energies of the Hamiltonian does not define a rational number, the system oscillates in time without a definite periodicity.
In particular, one notices that for $\kappa=0$ the relative phases $e^{-i \Delta \phi^{ab}_{n,s}}$ and $e^{-i \Delta \phi^{ac}_{n,s}}$ vanish (see Eqs.~(\ref{eqsC01A}) and (\ref{eqsC01B})), thus $P_{a \rightarrow c} = P_{a \rightarrow d} = 0$. The survivor probabilities $P_{b \rightarrow b}$, $P_{c \rightarrow c}$ and $P_{d \rightarrow d}$ have exactly the same value as the survivor probability of $\vert a (t) \rangle$, with an exception for the electric field interacting configuration where both tensor and pseudotensor couplings, $\kappa$ and $\mu$, do not vanish.
In this case, $P_{a \rightarrow a} = P_{b \rightarrow b} =P_{c \rightarrow c}$ are depicted in Fig.~\ref{eqsfig:03}.

Since Dirac bi-spinors are identified and quantified as {\em spin-parity} entangled states, the ionic states $\vert j \rangle$ shall also exhibit a profile of quantum entanglement.

The energy levels depicted in Fig.~\ref{eqsfig:level} and the {\em qubit} assignment from (\ref{eqsA08}) suggest the identification of two subsystems: the former one related to the total angular momentum quantum number, $\bm F$ ($\mathcal{S}_F$), and the latter one associated to the projection of the angular momentum onto the direction of the confining magnetic field, $\bm M$ ($\mathcal{S}_{M}$).
The energy levels are identified in agreement with the {\em qubit} assignment adopted in (\ref{eqsA08}).
Within such a framework, an internal ionic state $\vert j \rangle$ will evolve to a superposition between the four ionic states and shall exhibit quantum entanglement between $\mathcal{S}_F$ and $\mathcal{S}_{M_F}$ - which may be detected even from its departing configuration.
To quantify the entanglement along the time evolution of the quantum system, the ionic quantum state must be rewritten in terms of the oscillating ionic basis,
\begin{equation}
\vert \, j (t) \rangle = \displaystyle \sum_{k = a, \, ...,\, d} \bigg[ \sum_{(n,s) = 0,1} W^{j}_{n,s} M^{k}_{n,s} e^{- i \lambda_{n,s} t} \bigg] \vert k \rangle,
\end{equation}
such that the Bloch vector $\bm{a}_j (t) = \mbox{Tr}[\, \vert j(t) \rangle \langle j(t) \vert \, (\hat{I}_2^{(1)} \otimes \hat{\bm{\sigma}}^{(2)}) ]$ can be straightforwardly used to evaluate the quantum concurrence, $C[\rho]$, by means of (\ref{eqsB03}).
The \textit{spin-parity} entanglement is thus translated into the entanglement between the total angular momentum and its projection onto the direction of the trapping magnetic field.

\begin{figure}
\includegraphics[width = 16.5 cm]{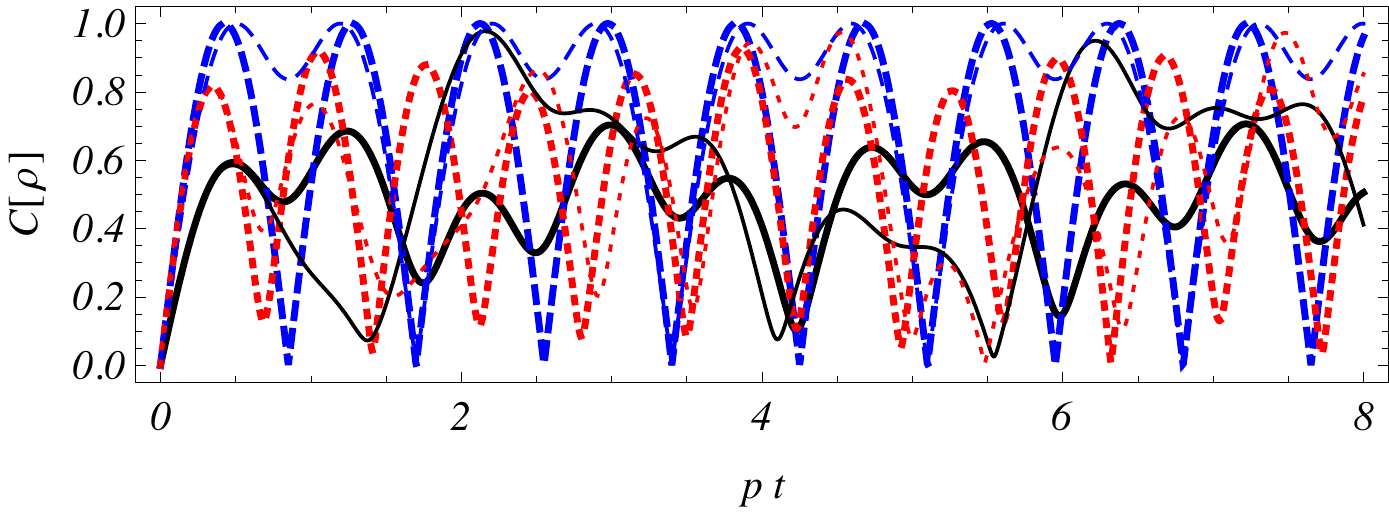}
\vspace{ .5cm}
\includegraphics[width = 16.5 cm]{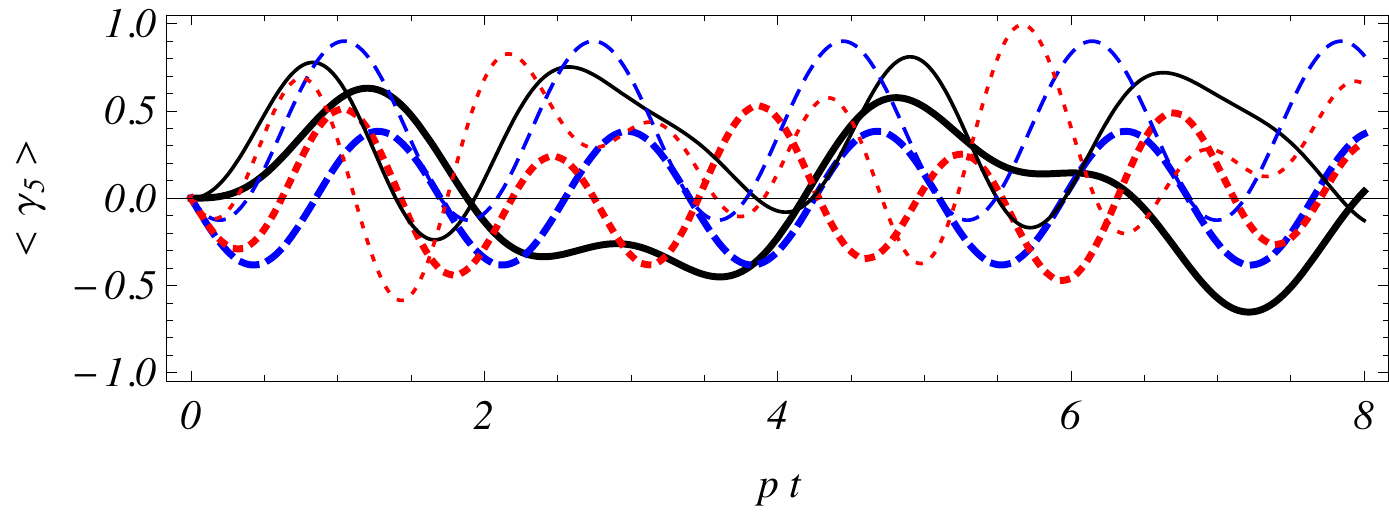}\quad
\caption{Quantum concurrence, $C[\rho]$, and averaged chirality, $\langle \hat{\gamma}_5 \rangle$, as functions of $p\,t$ for the same set of parameters considered in Fig.~\ref{eqsfig:02} (with the same correspondence to the plot-line styles). As the state is initially prepared like $\vert a \rangle$, for $t=0$, the state is separable.
Entanglement oscillates and then it vanishes for some particular values of $p \, t$. For $\kappa = 0$ the state is a superposition between $\vert a \rangle$ and $\vert d \rangle$, and the quantum concurrence varies between $0$ and $1$ (which indicates that the state is maximally entangled).
The averaged chirality also exhibits an oscillation pattern and it does not reach its maximum value at unity.
In this case, the quantum state contains a survival component given either by $(\vert a \rangle + \vert d \rangle)/\sqrt{2}$ or by $(\vert c \rangle + \vert b \rangle)/\sqrt{2}$.}
\label{eqsfig:04}
\end{figure}

Finally, the evaluation of the chiral operator $\hat{\gamma}_5$ is carried out in the bi-spinor basis results into
\begin{eqnarray}
\hat{\gamma}_5 &=& \displaystyle \sum_{(n,s) = 0,1} \sum_{(m,l) = 0,1} \big{[} \, W^a_{n,s} \big{(}\, W^d_{m,l} \, \big{)}^* + W^d_{n,s} \big{(}\, W^a_{m,l} \, \big{)}^* \nonumber \\
 &&\qquad\qquad\qquad +W^b_{n,s} \big{(}\, W^c_{m,l} \, \big{)}^* + W^c_{n,s} \big{(}\, W^b_{m,l} \, \big{)}^*\, \big{]} \, \vert \psi_{n,s} \rangle \langle \psi_{m,l} \vert,
\end{eqnarray}
and its average value is associated to measurements on superpositions between $\vert a \rangle$ and $\vert d \rangle$, and between $\vert b \rangle$ and $\vert c \rangle$ states (c. f. Eq. (\ref{eqsB06})).
The averaged chirality $\langle \hat{\gamma}_5 \rangle (t) = \mbox{Tr}[\hat{\gamma}_5 \, \vert a(t) \rangle \langle a(t) \vert]$ and the quantum concurrence $C[\rho(t)]$ are depicted in Fig.~\ref{eqsfig:04}.
Entanglement oscillates and vanishes for some specific values of $p \,t$.
For vanishing electric dipole moment, $\kappa = 0$ (dashed lines), the state is a superposition between $\vert a \rangle$ and $\vert d \rangle$, and its concurrence varies from zero, indicating a separable state, either $\vert a \rangle$ or $\vert d \rangle$, to unity, indicating that the state is the maximally entangled, $\vert \psi_{\mbox{max}} \rangle$, given by
\begin{equation}
\label{eqsC03}
\vert \psi_{\mbox{max}} \rangle = \frac{\vert a \rangle + e^{i \varphi} \, \vert d \rangle}{\sqrt{2}}.
\end{equation}

Differently from transition probabilities, concurrence has a well defined oscillation frequency, since, for its evaluation, one of the quantum system degrees of freedom was traced-out, leaving only the quantum state associated frequencies that are multiples one of each other. Otherwise, the averaged chirality also does not present a well defined oscillation pattern, vanishing for certain values of $p \, t$. It should be noticed that the averaged chirality does not attain its maximum value, such that the ionic state should have a component in $(\vert a \rangle + \vert d \rangle)/\sqrt{2}$ or in $(\vert c \rangle + \vert b \rangle)/\sqrt{2}$. When compared to the concurrence for electric dipole moment set equal to zero, the points where concurrence vanishes are exactly the same points where averaged chirality vanishes, since for such values of $p \, t$ the state is either $\vert a \rangle$ or $\vert d \rangle$, for which $P_{ad} = 1/2$ (c. f. Eq.~(\ref{eqsB06})). On the other hand, the extremum values of the averaged chirality correspond to the points for which $C[\rho] = 1$, as for this points the state has the form of (\ref{eqsC03}), for which $P_{ad}$ has an extremum.

The preparation and measurement of the setup discussed above can be accomplished by widely used experimental techniques. Once the ion vibrational ground state is prepared by laser cooling, the internal ionic state can be initialized via optical pumping with a probability larger than 99$\%$ \cite{exp01}. The optical pumping mechanism drives the atom up to reaching states inaccessible by the optical driver.
Circularly polarized light is used to pump the atom into one of its levels, and the initialization fidelity is limited by the quality of the driven laser polarization \cite{n012}. Detection of internal ionic states can be carried out by the electron shelving method which consists in detecting laser-induced fluorescence on an electric dipole allowed transition \cite{n012}.
This technique has been used, for instance, to measure the mean value of the position operator $\langle \hat{x} \rangle$ in a quantum simulation of the \textit{zitterbewegung} effect with trapped ions, by mapping the position state to the internal levels of the ion \cite{Nat06}.

Quantum concurrence cannot be directly measured, since that its definition is given through unphysical operations.
Although entanglement can be detected through some suitable properties of a given state \cite{exp02,exp03}, the dynamical evolution of the degree of entanglement requires a complete knowledge of the density matrix.
This can by achieved by quantum state tomography \cite{tom01,tom02}, i.e. the reconstruction of a density matrix by performing measurements on a large number of copies of the system. Quantum state tomography was performed for several configurations of trapped ions \cite{entjc01,exp05,exp06,exp07,exp08,exp09,exp10,exp11}.
It has included the measurement of ion motion states \cite{exp07,exp08}, detection of multipartite entanglement \cite{exp10} and characterization of entanglement for quantum computation and quantum information purposes \cite{exp09,exp11}. Although the quantum state tomography allows for reconstructing all density matrix elements, additional/partial information of state dynamics, as required for the characterization of the evaluation of average chirality, might be accomplished by quantum simulation analogous to that used for measuring the mean value of the position operator \cite{Nat06,exp12,exp13}.

\section{Conclusions}

Single trapped ions with quantum states driven by Jaynes-Cummings, anti-Jaynes-Cummings and carrier interactions have been treated as a suitable experimental platform for simulating Dirac-like dynamics, in particular, for Dirac Hamiltonians which include external electrostatic potentials.
This simulation method has been used to identify and measure relativistic-like quantum effects that, in high energy physics, are shown to be inaccessible by the current experimental apparatus.
After revealing and explaining how some features, which are typical from bi-spinor Dirac-like systems, are related to trapped ion physics, some engendered ion configurations have been prepared for supporting, for example, the detection of local quantum correlations \cite{Nat01}, the simulation of open quantum systems \cite{Nat02,Nat04}, the construction of microwave quantum logic gates \cite{Nat03}, and the prediction of the existence of mesoscopic cat states \cite{n013} as well as quantum phase transitions \cite{Nat05,n014}.

Incremental issues related to the such features, in particular, those ones related to the computation of quantum transition probabilities and quantum entanglement, have been addressed in this manuscript.
The scenario considered here corresponds to a Dirac-like system simulated by ion traps for which the accessible experimental driving variables are mapped into a relativistic system for Dirac particles non-minimally coupling to an electrostatic field.
Single and coupled effects of Dirac-like tensor and pseudotensor potentials, as well as the influence of mass-like parameter of the relativistic bi-spinor, have been considered at an analysis which was addressed to an one-dimensional bi-spinor particle propagation in the $x$ direction, with the electrostatic field lying in the $x-y$ plane.
Once mapped onto trapped ion states, four level system transition probabilities due to typical quantum oscillations have been analytically obtained, and an oscillation pattern similar to those involving two and three level systems \cite{Neu01,Neu02} has been identified.

By interpreting each ionic state as a quantum superposition of two-{\em qubit} states, i. e. one {\em qubit} associated to the total angular momentum and the another one associated to the projection of the angular momentum onto the direction of the trapping magnetic field, the associated quantum entanglement has been quantified by means of the quantum concurrence.
The results indicate that quantum entanglement measurements should exhibit a time-dependent oscillating pattern, such that it only attains a maximum value in the absence of electric dipole coupling, i. e. when $\kappa=0$.
It has been noticed that an internal ionic state oscillates between $\vert a \rangle$ and $\vert d \rangle$ states, being maximally entangled for the state corresponding to a quantum superposition written as $(\vert a \rangle + e^{i \varphi} \vert d \rangle) /\sqrt{2}$.
In addition, our results indicate that carrier interactions associated to $\kappa$ actively drive the suppression of the quantum entanglement.

Likewise, the averaged chirality is tested as a quantifier of the maximal superposition between $\vert a\rangle$ and $\vert d\rangle$, and between $\vert c\rangle$ and $\vert b\rangle$ ionic levels.
In particular, for the case of a vanishing dipole moment, $\kappa=0$, the modulus of the averaged chirality and the quantum concurrence are concomitantly null as well as they have coincident maximal point values.
Such a quantum correlational correspondence between averaged chirality and quantum concurrence exhibits a similar connection to that between the time-reversal quantum operator and quantum entanglement, a point which deserves some subsequent investigation.

To end up, measuring relativistic effects in tabletop experiments has been identified as one of the main purposes of the quantum simulation of Dirac equation \cite{Nat06,n001,n002,n003,n004,n005}.
For the Dirac dynamics which includes tensor and pseudotensor potentials (c. f. Eq.~(\ref{eqsA00})), single trapped ion platforms work as to measure, for instance, spin precession and degeneracy lifting in connection with $CP$ violation (present at supersymmetric models) \cite{n005}. 
A first step towards more complex quantum simulations of Dirac-like systems has been given by Ref.~\cite{Nat06}.
Those outstanding results suggest that the entanglement structure of Dirac bi-spinors \cite{n008,n010} can be probed via trapped ions even if experimental techniques are still underestimated. 
In that case, for the trapped ion platforms, the only observable that can straightforwardly be measured by fluorescence techniques is $\hat{\sigma}_z$ (c. f. Eqs~(\ref{eqsA01})-(\ref{eqsA04})). Otherwise, extra laser pulses can be used to map other observables onto $\hat\sigma_z$. 
As to determine the averaged values which are relevant for computing quantum concurrence and chirality, a novel state-dependent displacement operation has to be engendered, in order to connect such (theoretical) averaged values with phenomenologically detectable measures of $\hat{\sigma}_z$.
From the theoretical construction, the quantum entanglement encoded by the solutions of Dirac equation can be simulated even when the ion is prepared in one of its internal levels. Despite following the above statements, the connection of such entanglement quantifier observables with the measurement techniques seems to be not so trivial and deserves a more careful investigation.
Furthermore, environment effects \cite{n020} might be coupled to the dynamics and simulated by JC interactions \cite{conc01,conc02} in order to include, for instance, decoherence effects.

Finally, the Dirac equation with external fields also describes low energy excitation of mono and bi-layer graphene with imperfections \cite{conc03}, such that the formalism and procedures here discussed can also be employed for a complete characterization of such excitations, which includes the computation of survivor probabilities and electron-electron or electron-hole entanglement quantifiers.

{\em Acknowledgments - The work of AEB is supported by the Brazilian Agencies FAPESP (grant 15/05903-4) and CNPq (grant 300809/2013-1). The work of VASVB is supported by the Brazilian Agency CNPq (grant 140900/2014-4).}

\section*{Appendix}

In terms of Lie algebras and Lie groups, the representations of $sl(2,\mathbb{C})\oplus sl(2,\mathbb{C})$ algebra, which correspond to the Lie algebra of the $SL(2,\mathbb{C})\otimes SL(2,\mathbb{C})$ group, are irreducible, i.e. these representations correspond to tensor products between linear complex representations of $sl(2,\mathbb{C})$, as it is observed by considering the restriction to the subgroup $SU(2)\otimes SU(2) \subset SL(2,\mathbb{C})\otimes SL(2,\mathbb{C})$. Unitary irreducible representations of $SU(2)\otimes SU(2)$ are precisely tensor products between unitary representations of $SU(2)$. These representations establish a one-to-one correspondence with the group $SL(2,\mathbb{C})\otimes SL(2,\mathbb{C})$, and considering that it is a simply connected group, a one-to-one correspondence with the algebra $sl(2,\mathbb{C})\oplus sl(2,\mathbb{C})$.

The existence of {\em inequivalent representations} of $SU(2) \otimes SU(2)$ follows from the above mentioned one-to-one correspondences. Such representations may not correspond to all the representations of $SL(2,\mathbb{C})\otimes SL(2,\mathbb{C})$ (therefore, of the proper Lorentz transformations that compose the $SO(3,1)$ group), instead they describe a subset of $SO(4) \equiv SO(3)\otimes SO(3)$ transformations, for instance, those which include the group of double covering rotations.

As the transformations of $SU(2)\otimes SU(2)$ can be described by a subset of $SL(2,\mathbb{C})\otimes SL(2,\mathbb{C})$, one may choose at least two {\em inequivalent} subsets of $SU(2)$ generators, such that $SU(2)\otimes SU(2) \subset SL(2,\mathbb{C})\otimes SL(2,\mathbb{C})$, with each generator having its own irreducible representations ($irrep$) simbolicaly described by $irrep (su_{\xi}(2)\oplus su_{\chi}(2))$. Therefore, a {\em spinor} $\xi$ described by $(\frac{1}{2},\,0)$ transforms as a {\em doublet} - object of the fundamental representation - of $SU_{\xi}(2)$, and as a singlet - object ``transparent'' to transformations - of the $SU_{\chi} (2)$ group. By adopting the notation $(\mbf{dim}(SU_{\xi}(2)),\mbf{dim}(SU_{\chi}(2)))$, the {\em spinor} $\xi$ is an object given by (\mbf{2},\mbf{1}). Following the same idea, the {\em spinor} $\chi$, described by $(0,\,\frac{1}{2})$, transforms as a {\em singlet} of $SU_{\xi}(2)$ and as a {\em doublet} of $SU_{\chi}(2)$.

With respect to the representations of $SL(2,\mathbb{C})$ one has the objects:

$(\mbf{1},\mbf{1})$ - a {\em scalar} or {\em sinlget}, with angular momentum projection $j = 0$;

$(\mbf{2},\mbf{1})$ - a {\em spinor} $(\frac{1}{2},\,0)$, commonly referred as {\em left-handed}, with angular momentum projection $j = 1/2$;

$(\mbf{1},\mbf{2})$ - a {\em spinor} $(0,\,\frac{1}{2})$, commonly referred as {\em right-handed}, with angular momentum projection $j = 1/2$;

$(\mbf{2},\mbf{2})$ - a {\em vector} or {\em doublet}, with angular momentum projection $j = 0$ and $j = 1$;

The fundamental objects of an {\em irrep} can be used to construct more complex objects. With respect to the representations of $SL(2,\mathbb{C})$ one may construct, for example,
$ (\mbf{1},\mbf{2}) \otimes (\mbf{1},\mbf{2}) \equiv (\mbf{1},\mbf{1}) \oplus (\mbf{1},\mbf{3}),$,
a representation that composes Lorentz tensors like
\begin{equation}
C_{\alpha\beta}\bb{\mt{x}} = \epsilon_{\alpha\beta} D\bb{\mt{x}} + G_{\alpha\beta}\bb{\mt{x}},
\end{equation}
where $D\bb{\mt{x}}$ is a scalar, and $G_{\alpha\beta} = G_{\beta\alpha}$ is totally symmetric, or even
$ (\mbf{2},\mbf{1}) \otimes (\mbf{1},\mbf{2}) \equiv (\mbf{2},\mbf{2})$,
such that
$ (\mbf{2},\mbf{2}) \otimes (\mbf{2},\mbf{2}) \equiv (\mbf{1},\mbf{1}) \oplus (\mbf{1},\mbf{3}) \oplus (\mbf{3},\mbf{1}) \oplus (\mbf{3},\mbf{3})
$,
that composes Lorentz tensors like
\begin{equation}
\varphi^{\mu\nu}\bb{\mt{x}} = A^{\mu\nu}\bb{\mt{x}} + S^{\mu\nu}\bb{\mt{x}} + \frac{1}{4}g^{\mu\nu} \Theta\bb{\mt{x}},
\end{equation}
which correspond to a decomposition into smaller {\em irreps} related to the Poincaré classes quoted at \cite{n009}, with $A^{\mu\nu} \equiv (\mbf{1},\mbf{3}) \oplus (\mbf{3},\mbf{1})$ totally anti-symmetric by the index interchange $\mu\leftrightarrow \nu$, $S^{\mu\nu}\equiv (\mbf{3},\mbf{3})$ totally symmetric by the index interchange $\mu\leftrightarrow \nu$, and $\Theta \equiv (\mbf{1},\mbf{1})$ as a Lorentz scalar, which is multiplied by the metric tensor, $g^{\mu\nu}$.

The above properties support the construction of the Dirac Hamiltonian dynamics through a group representation described by a direct product between two algebras which compose a subset of the group $SL(2,\mathbb{C})\otimes SL(2,\mathbb{C})$, the group $SU(2)\otimes SU(2)$. Majorana, Weyl and some additional classes of spinor equations can also be driven by other subsets of $SL(2,\mathbb{C})\otimes SL(2,\mathbb{C})$.

In quantum mechanics, the free particle Dirac Hamiltonian reads
\begin{equation}
\hat{H}_{D}=\hat{\mbf{\alpha}}\cdot \hat{\mbf{p}}+m \hat{\beta},
\label{eqshamdirac}
\end{equation}
where the Dirac operators, $\hat{\mbox{\boldmath$\alpha$}}$ and $\hat{\beta}$, have already been identified by Eq.~(\ref{eqsAAA}) (now given in natural units, $c = \hbar = 1$).
For the corresponding state vectors, one writes
$ \psi^{\dagger} \bb{\mt{x}} =\left(
\psi^{\dagger}_{\L}\bb{\mt{x}},
\psi^{\dagger}_{\R}\bb{\mt{x}}
\right)\equiv (\mbf{2},\mbf{2})$,
with {\em right-handed} and {\em left-handed} {\em spinors},
\begin{equation}
(\mbf{2},\mbf{1}) \equiv \psi _{\L}\bb{\mt{x}} =\left(
\begin{array}{c}
\psi _{\L\1}\bb{\mt{x}} \\
\psi _{\L\2}\bb{\mt{x}}
\end{array}
\right) ,\qquad (\mbf{1},\mbf{2}) \equiv\psi _{\R}\left( t\right) =\left(
\begin{array}{c}
\psi _{\R\1}\bb{\mt{x}} \\
\psi _{\R\2}\bb{\mt{x}}
\end{array}
\right).
\end{equation}
The free particle Dirac equation is thus mapped by two coupled differential equations for the $\psi_{\L}\bb{\mt{x}}$ e $\psi _{\R}\bb{\mt{x}}$, respectively,
\begin{eqnarray*}
i{\overline{\sigma}}^{\mu }\partial _{\mu }\psi _{\L}\bb{\mt{x}}
-m\psi _{\R}\bb{\mt{x}} &=&0, \\
i{\sigma }^{\mu }\partial _{\mu }\psi _{\R}\bb{\mt{x}} -m\psi
_{\L}\bb{\mt{x}} &=&0,
\end{eqnarray*}
where, in the {\em chiral representation}, $\hat{\mathbb{I}}_{\2} \otimes \hat{\mbf{\boldsymbol{\sigma }}} =
\sigma^{\mu}$ and $\hat{\mathbb{I}}_{\2} \otimes(-\hat{\mbf{\boldsymbol{\sigma }}}) ={\overline{\sigma}}^{\mu}$, for which the Lagrangian density reads
\begin{equation}
\mathcal{L}=i\psi _{\L}^{\dagger }{\overline{\sigma}}^{\mu }\partial
_{\mu }\psi _{\L}+i\psi _{\R}^{\dagger }\mathbf{\sigma }^{\mu }\partial _{\mu
}\psi _{\R}-m\left( \psi _{\L}^{\dagger }\psi _{\R}+\psi _{\R}^{\dagger }\psi
_{\L}\right),
\end{equation}
from which a correspondence with the {\em spinor} helicity is identified.

An alternative interpretation for the spinors is obtained when Dirac equation is written in terms of Kronecker products between Pauli matrices. By the interpretation of quantum mechanics as a special information theory for particles and fields, one can identify the Dirac equation solutions as it was described by two {\em qubits} states encoded in a massive particle whose dynamics is represented by continuous variables, which can be the linear momentum or the position. The $\hat{\bm{\alpha}}$ and $\hat{\beta}$ matrices written in terms of Pauli matrices reads \cite{n008}
\begin{equation}
\quad \hat{\mbf{\boldsymbol{
\ \alpha }}}= \hat{\sigma} _{x}^{\left( 1\right) }\otimes
 \hat{\mbf{\boldsymbol{\sigma }}}^{\left( 2\right) } ,~
\text{and}~~\hat{\beta}=\hat{\sigma}_{z}^{\left( 1\right)}\otimes \hat{\mathbb{I}}_{\2}^{\left( 2\right)},
\end{equation}
with the subscripts $1$ and $2$ referring to the {\em qubits} $1$ and $2$.

Within this framework, the $\mbox{SU}(2)\otimes \mbox{SU}(2)$ representation of Dirac bi-spinors is generated by the free Hamiltonian given in terms of two-{\em qubit} operators, ${H}_{D}={\hat{\sigma}}_{x}^{\left( 1\right) }\otimes \left(
{\bm p}\cdot {{\hat{\mbox{\boldmath$\sigma$}}}}^{\left( 2\right) }\right) +m \,
{\hat{\sigma} }_{z}^{\left( 1\right)}\otimes {I}^{(2)}_{2}$, for which the eigenstates are written in terms of a sum of direct products describing \textit{spin-parity} entangled states,
\begin{eqnarray}
\label{eqsB02}
\lefteqn{\left\vert \Psi ^{s}({\bm p},\,t)\right\rangle=
e^{i(-1)^{s}\,E_{p}\,t}\left\vert \psi ^{s}({\bm p})\right\rangle
}\\
&&= e^{i(-1)^{s}\,E_{p}\,t} N_{s}\left( p\right) \notag
 \left[ \left\vert
+\right\rangle _{1}\otimes \left\vert u({\bm p})\right\rangle _{2}+\left(
\frac{p}{E_{p}+(-1)^{s+1}m}\right) |-\rangle _{1}\,\otimes \left( {\bm p}
\cdot {\hat{\mbox{\boldmath$\sigma$} }}^{\left( 2\right) }\left\vert u(\bm{p}
)\right\rangle _{2}\right) \right],
\end{eqnarray}
where $s = 0,\, 1$ stands for particle/antiparticle associated frequencies, and the spinor character is given by $\left\vert u(\mbf{p})\right\rangle _s$ \cite{n008,n009}.
The state $\left\vert u(\mbf{p})\right\rangle _{\2}$ is a {\em bi-spinor} that describes the dynamics of a fermion (in momentum representation) coupled to its {\em spin}, which describes a magnetic dipole moment in the case of a coupling with an external magnetic field. The state (\ref{eqsB02}) is a superposition between parity eigenstates and therefore it does not exhibit a defined intrinsic parity.
For the {\em qubit} $1$, the {\em kets} $\left\vert + \right\rangle _{\1}$ e $ \vert - \rangle _{\1}$ are identified as the intrinsic parity eigenstates of the fermion.
These states are orthogonal, $\left\langle \pm |\pm (\mp )\right\rangle _{\1}=1(0)$, and one has $\left\langle \psi ^{s}(\mbf{p},\,t)|\psi ^{s}(\mbf{p},\,t)\right\rangle =\left\langle u(\mbf{p})|u(\mbf{p})\right\rangle _{\2}$, where the normalization is given by
\begin{equation}
N_{s}(p)=\frac{1}{\sqrt{2}}\left( 1+(-1)^{s+1}\frac{m}{E_{p}}\right) ^{\1/\2},
\label{eqsnorm}
\end{equation}
such that the local probability distribution for the momenta is normalized by $\int {\mt{d}^{\3}\hspace{-.07 cm}\mbf{p}\, \left\langle u(\mbf{p})|u(\mbf{p})\right\rangle}_{\2}=1$. Thus, one notices that {\em spinors} and Dirac matrices represent the direct product between the internal degrees of freedom associated to a {\em spin} $1/2$ massive fermion, parameterized by its linear momentum.

The total parity operator $\hat{P}$ acts on the direct product $\left\vert \pm \right\rangle _{\1}\otimes \left\vert u(\mbf{p})\right\rangle_{\2}$ in the form of $$\hat{P}\left( \left\vert \pm \right\rangle _{\1}\otimes \left\vert u(\mbf{p})\right\rangle _{\2}\right) =\pm \left( \left\vert \pm \right\rangle_{\1}\otimes \left\vert u(-\mbf{p})\right\rangle _{\2}\right),$$ and, for instance, it corresponds to the product of two operators: intrinsic parity, $\hat{P}^{int}$ (with two eigenvalues, $\hat{P}^{int}\left\vert \pm \right\rangle =\pm \left\vert\pm \right\rangle $) and spatial parity $\hat{P}^{s}$ (with $\hat{P}^{s}\varphi \left( \mbf{p}\right) =\varphi \left( -\mbf{p}\right) $). By applying $\hat{P}^{int}=\hat{\beta} =\hat{\sigma} _{z}^{\left( 1\right)}\otimes \hat{\mathbb{I}}_{\2}^{\left( 2\right) }$ to $\left\vert \psi ^{s}(\mbf{p},\,t)\right\rangle $, following Eq.~(\ref{eqsB02}), it follows that $ \hat{P} ^{\mi\1}=\hat{P}$, and the spatial parity resembles $\hat{P}^{int}$, as well as $\left( \hat{P}^{int}\right)^{\2}=\hat{\mathbb{I}}_{\2}^{\left( 1\right) }\otimes \hat{\mathbb{I}}_{\2}^{\left( 2\right) }$,
\begin{equation}
\hat{P}^{s}\left\{
\begin{array}{c}
\mbf{x} \\
\mbf{p}
\end{array}
\right\} \hat{P}^{s}=-\left\{
\begin{array}{c}
\mbf{x} \\
\mbf{p}
\end{array}
\right\} \text{,\quad and \quad}\hat{P}^{s}\left\{
\begin{array}{c}
\mbf{l} \\
\mbf{\sigma}
\end{array}
\right\} \hat{P}^{s}= + \left\{
\begin{array}{c}
\mbf{l} \\
\mbf{\sigma}
\end{array}
\right\} ,
\end{equation}
where the $+$ and $-$ signals are relative to axial and polar vectors, respectively.

\newpage

\begin{thebibliography}{99}
\bibitem{Nat01}
M. Gessner, M. Ramm, T. Pruttivarasin, A. Buchleitner, H.-P. Breuer and H. Haeffner, Nature Physics {\bf 10}, 105 (2014).
\bibitem{Nat02}
P. Schindler, M. M\"uller, D. Nigg, J. T. Barreiro, E. A. Martinez, M. Hennrich, T. Monz, S. Diehl, P. Zoller and R. Blatt, Nature Physics {\bf 9}, 361 (2013).
\bibitem{Nat03}
C. Ospelkaus, U. Warring, Y. Colombe, K. R. Brown, J. M. Amini, D. Leibfried and D. J. Wineland, Nature {\bf 476}, 181 (2011).
\bibitem{Nat04}
Julio T. Barreiro, Markus M\"uller, Philipp Schindler, Daniel Nigg, Thomas Monz, Michael Chwalla, Markus Hennrich, Christian F. Roos, Peter Zoller and Rainer Blatt, Nature {\bf 470}, 486 (2011).
\bibitem{Nat05}
R. Islam, E. E. Edwards, K. Kim, S. Korenblit, C. Noh, H. Carmichael, G.-D.Lin, L.-M. Duan, C.-C. Joseph Wang, J. K. Freericks and C. Monroe, Nature Commun. {\bf 2}, 377 (2011).
\bibitem{n001}
L. Lamata, J. Le\'{o}n, T. Sch\"{a}tz, and E. Solano, Phys. Rev. Lett. {\bf 98}, 253005 (2007). 
\bibitem{n002}
J. Casanova, J. J. Garc\'{\i}a-Ripoll, R. Gerritsma, C. F. Roos and E. Solano, Phys. Rev. {\bf A82}, 020101(R) (2010). 
\bibitem{n003}
R. Gerritsma, B. P. Lanyon, G. Kirchmair, F. Z\"ahringer, C. Hempel, J. Casanova, J. J. Garc\'ia-Ripoll, E. Solano, R. Blatt, and C. F. Roos, Phys. Rev. Lett. {\bf 106}, 060503 (2011). 
\bibitem{Nat06}
R. Gerritsma, G. Kirchmais, F. Zahringer, E. Solando, R. Blatt and C. F. Ross, Nature {\bf 463} 68 (2010). 
\bibitem{n004}
A. Bermudez, M. A. Martin-Delgado and E. Solano, Phys. Rev. {\bf A76}, 041801(R) (2007). 
\bibitem{n005}
T. G. Tenev, P. A. Ivanov and N. V. Vitanov, Phys. Rev. {\bf A87}, 022103 (2013). 
\bibitem{n006}
L. Lamata, J. Casanova, R. Gerritsma, C. F. Roos, J. J. Garc\'ia-Ripoll and E. Solano, New Journal of Physics {\bf 13}, 095003 (2011). 
\bibitem{new01}
T. E. Lee, U. Alvarez-Rodriguez, X.-H. Cheng, L. Lamata and E. Solano, Phys. Rev. {\bf A92}, 032129 (2015). 
\bibitem{n008}
A. E. Bernardini and S. S. Mizrahi, Phys. Scr. {\bf 89}, 075105 (2014). 
\bibitem{n009}
V. A. S. V. Bittencourt and A. E. Bernardini, Annals of Physics {\bf 364}, 182 (2016). 
\bibitem{entjc01}
H. H\"affner, C. F. Roos and R. Blatt, Phys. Rep. {\bf 429}, 155 (2008). 
\bibitem{entjc04}
C. Monroe, D. M. Meekhof, B. E. King and D. J. Wineland, Science {\bf 22}, 1131 (1996).
\bibitem{entjc02}
D. Ito, K. Moriand and E. Carriere, Nuovo Cimento A {\bf 51}, 1119 (1967).
\bibitem{entjc03}
P. Rozmej and R. Arvieu, J. Phys. A: Math. Gen. {\bf 32}, 5367 (1999).
\bibitem{n014}
A. Bermudez, M. A. Matin-Delgado and A. Luis, Phys. Rev. A {\bf 77}, 063815 (2008). 
\bibitem{n010}
V. A. S. V. Bittencourt, S. S. Mizrahi, and A. E. Bernardini, Annals of Physics {\bf 355}, 35 (2015).
\bibitem{n011}
A. E. Bernardini, J. Phys. G: Nucl. Part. Phys. {\bf 32}, 9 (2006).
\bibitem{n011A}
A. E. Bernardini, J. Phys. A: Math. Gen. {\bf 39}, 7089 (2006).
\bibitem{n011B}
A. E. Bernardini, Eur. Phys. J. {\bf C 50}, 673 (2007).
\bibitem{n012}
D. Liebfried, R. Blatt, C. Monroe and D. Wineland, Rev. Mod. Phys. {\bf 75}, 281 (2003). 
\bibitem{n015}
B. Thaller, \textit{The Dirac Equation} (Springer-Verlag, New York, 1992).
\bibitem{n018}
A. Peres and D. R. Terno, Rev. Mod. Phys. {\bf 76}, 93 (2004). 
\bibitem{n019}
A. Peres, Phys. Rev. Lett. {\bf 77}, 8 (1996). 
\bibitem{n020}
H. P. Breuer, F. Petruccione, {\em The Theory of Open Quantum Systems} (Oxford University Press, New York, 2002).
\bibitem{n021}
W. K. Wootters, Phys. Rev. Lett. {\bf 80}, 2245 (1998).
\bibitem{exp01}
D. J. Wineland, J. C. Bergquist, W. M. Itano and R. E. Drullinger, Op. Lett. {\bf 5}, 6 (1980). 
\bibitem{exp02}
C. A. Sackett, D. Kielpinski, B. E. King, C. Langer, V. Meyer, C. J.Myatt, M. Rowe, Q. A. Turchette, W. M. Itano, D. J. Wineland and C. Monroe, Nature {\bf 404}, 256 (2000). 
\bibitem{exp03}
C. J. Myatt, B. E. King, Q. A. Turchette, C. A. Sackett, D. Kielpinski, W. M. Itano, C. Monroe and D. J. Wineland, Nature {\bf 403}, 269 (2000). 
\bibitem{tom01}
U. Leonhardt, Phys. Rev. Lett. {\bf 74}, 4101 (1995). 
\bibitem{tom02}
Z. Hradil, Phys. Rev. A {\bf 55}, R1561 (1997). 
\bibitem{exp05}
C. F. Roos, G. P. Lancaster, M. Riebe, H. H\"affner, W. H\"ansel, S. Hulde, C. Becher, J. Eschner, F. Schmidt-Kaler and R. Blatt, Phys. Rev. Lett. {\bf 92}, 22 (2004). 
\bibitem{exp06}
H. H\"affner, F. Schmidt-Kaler, W. H\"ansel, C. F. Roos, T. K\"orber, M. Chwalla, M. Riebe, J. Benhelm, U. D. Rapol, C. Becher and R. Blatt, Appl. Phys. B {\bf 81}, 151 (2005). 
\bibitem{exp07}
D. Leibfried, D. M. Meekhof, B. E. King, C. Monroe, W. M. Itano and D. J. Wineland, Phys. Rev. Lett. {\bf 77} 4281 (1996). 
\bibitem{exp08}
J. F. Poyatos, R. Walser, J. I. Cirac, P. Zoller and R. Blatt, Phys. Rev. A {\bf 53} R1966 (1996). 
\bibitem{exp09}
M. Riebe, K. Kim, P. Schindler, T. Monz, P. O. Schmidt, T. K. K\"orber, W. H\"ansel, H. H\"affner, C. F. Roos and R. Blatt, Phys. Rev. Lett. {\bf 97}, 220407 (2006). 
\bibitem{exp10}
H. H\"affner, W. H\"ansel, C. F. Roos, J. Benhelm, D. Chek-al-kar, M. Chwalla, T. K\"orber, U. D. Rapol, M. Riebe, P. O. Schmidt, C. Becher, O. G\"uhne, W. D\"ur and R. Blatt, Nature {\bf 438}, 643-646 (2005). 
\bibitem{exp11}
C. J. Ballance, V. M. Schaefer, J. P. Home, D. J. Szwer, S. C. Webster, D. T. C. Allcock, N. M. Linke, T. P. Harty, D. P. L. Aude Craik, D. N Stacey, A. M. Steane and D. M. Lucas, Nature {\bf 528}, 384 (2015). 
\bibitem{exp12}
M. Mohseni and D. A. Lidar, Phys. Rev. Lett. {\bf 97}, 170501 (2006). 
\bibitem{exp13}
M. Mohseni, A. T. Rezakhani and D. A. Lidar, Phys. Rev. A {\bf 77}, 032322 (2008). 
\bibitem{n013}
A. Bermudez, M. A. Martin-Delgado and E. Solano, Phys. Rev. Lett. {\bf 99}, 123602 (2007). 
\bibitem{Neu01}
A. E. Bernardini and V. A. S. V. Bittencourt, Astropart. Phys. {\bf 41}, 31 (2013).
\bibitem{Neu02}
V. A. S. V. Bittencourt, C. J. Villas-Boas and A. E. Bernardini, EuroPhysics Lett. {\bf 41 }, 31 (2014).
\bibitem{conc01}
T. Yu and J. H. Eberly, Phys. Rev. Lett. {\bf 93}, 14040 (2004).
\bibitem{conc02}
T. Yu and J. H. Eberly, Opt. Comm. {\bf 264}, 393 (2006).
\bibitem{conc03}
N. M. R. Peres, F. Guinea, and A. H. Castro Neto, Phys. Rev. B {\bf 73}, 125411 (2006). 
\end{thebibliography}
\end{document}